\documentclass[aps,pra,twocolumn,superscriptaddress,nofootinbib,10pt]{revtex4-1}
\usepackage[english]{babel}
\usepackage{amsmath}
\usepackage{url}
\usepackage{graphicx, subfigure}
\usepackage[latin1]{inputenc}
\newcommand{\vt}[1]{\ensuremath{\boldsymbol{#1}}}
\usepackage{amssymb}
\usepackage{color}
\newcommand{\blue}[1]{\textcolor{blue}{#1}}

\RequirePackage[top=2cm]{geometry}
\begin{document}

\title{Error estimates for solid-state density-functional theory predictions: an overview by means of the ground-state elemental crystals}

\author{K. Lejaeghere}
\author{V. Van Speybroeck}
\affiliation{Center for Molecular Modeling, Ghent University, Technologiepark 903, BE-9052 Zwijnaarde, Belgium}

\author{G. Van Oost}
\affiliation{Department of Applied Physics, Ghent University, Sint-Pietersnieuwstraat 41, BE-9000 Ghent, Belgium}

\author{S. Cottenier}
\email[]{Stefaan.Cottenier@UGent.be}
\affiliation{Center for Molecular Modeling, Ghent University, Technologiepark 903, BE-9052 Zwijnaarde, Belgium}
\affiliation{Department of Materials Science and Engineering, Ghent University, Technologiepark 903, BE-9052 Zwijnaarde, Belgium}

\begin{abstract}
\begin{center}
\blue{\texttt{This is an Author's Accepted Manuscript of an article published in \emph{Critical Reviews in Solid State and Materials Sciences} 39, p 1 (2014), available online at \url{http://www.tandfonline.com/bsms} and with the DOI: 10.1080/10408436.2013.772503.}}
\end{center}

Predictions of observable properties by density-functional theory calculations (DFT) are used increasingly often by experimental condensed-matter physicists and materials engineers as data. These predictions are used to analyze recent measurements, or to plan future experiments in a rational way. Increasingly more experimental scientists in these fields therefore face the natural question: what is the expected error for such a first-principles prediction? Information and experience about this question is implicitly available in the computational community, scattered over two decades of literature. The present review aims to summarize and quantify this implicit knowledge. This eventually leads to a practical protocol that allows any scientist -- experimental or theoretical -- to determine justifiable error estimates for many basic property predictions, without having to perform additional DFT calculations.

A central role is played by a large and diverse test set of crystalline solids, containing all ground-state elemental crystals (except most lanthanides). For several
properties of each crystal, the difference between DFT results and experimental values is assessed. We discuss trends in these deviations and review explanations suggested in the literature.

A prerequisite for such an error analysis is that different implementations of the same first-principles formalism provide the same predictions. Therefore, the reproducibility of predictions across several mainstream methods and codes is discussed too. A quality factor $\Delta$ expresses the spread in predictions from two distinct DFT implementations by a single number. To compare the PAW method to the highly accurate APW+lo approach, a code assessment of VASP and GPAW (PAW) with respect to WIEN2k (APW+lo) yields $\Delta$-values of 1.9 and 3.3\,meV/atom, respectively. In both cases the PAW potentials recommended by the respective codes have been used.
These differences are an order of magnitude smaller than the typical difference with experiment, and therefore predictions by APW+lo and PAW are for practical purposes identical.
\end{abstract}

\maketitle

\tableofcontents

\section{Introduction}

Density-functional theory \cite{HohenbergKohn, KohnSham} (DFT) remains one of the most popular methods to treat both model systems and realistic materials in a quantum mechanical way \cite{Segallrev, Liurev, Martinrev, Capelle, Hafnerrev, vasp2}. In condensed-matter physics, DFT is not only used to understand the observed behavior of solids, but increasingly more to predict characteristics of compounds that have not yet been determined experimentally (for example in Refs.~\onlinecite{Ma, Hautier2, Jain}). In both cases the first-principles results provide a point of reference, either to analyze data from measurements or to plan future experiments. It is therefore essential to have a quantitative idea of the expected deviation between a DFT prediction of a certain property and the corresponding experimental value. Error estimates are routinely provided in experimental physics, but in DFT applications this is much less common practice. When confronted with a disagreement between theory and experiment, 
one 
usually resorts to higher-order levels of theory instead \cite{Schimka, Gaston2010, Bil}.

DFT as such is an exact reformulation of quantum physics, and does not involve any approximation. From a purely theoretical point of view, it should lead to exact predictions, with no need for an error estimate. In practice, however, one requires an educated guess for an essential ingredient of DFT: the exchange-correlation functional 
(hereafter referred to as `functional'). Apart from this main approximation, there are some other features that go beyond DFT in the way it is usually applied, such as the failure of the Born-Oppenheimer approximation \cite{failureBO, failureBO2} or high-$Z$ radiative and other corrections from quantum electrodynamics \cite{Pyykko, qed1, qed2}. They affect the results to some extent as well, but they are generally much less important than the particular choice of the exchange-correlation functional.
For any of these choices, DFT predictions will not agree perfectly with experimental observations. Deviations of this kind will be referred to as the \emph{intrinsic error} for a particular functional.

Regardless of the difference with experiment, however, 
all predictions should be independent of how the DFT (Kohn-Sham) equations are solved numerically: each of the many 
available DFT implementations should give identical results for the same functional. In reality, 
there will be some 
scatter in the predictions of different codes \cite{g2-1,Kiejna}, as each of them introduces a distinct amount of numerical noise. This second source 
of fluctuations leads to a \emph{numerical error}. It is therefore also important to assess to what extent predicted properties vary among DFT approaches.

The goal of this work is to quantify the knowledge about these two kinds of DFT errors, intrinsic and numerical ones, and --- where relevant --- to review the physics behind them.

A way to obtain insight into computational errors is by means of benchmark studies, examining the performance of different implementations and functionals for a large set of materials and properties. For molecular benchmark sets the intrinsic errors have already been assessed in great detail, often with the aim of selecting the best functional for a particular property (for example in Refs.~\onlinecite{PBE, Kotochigova, Kotochigovaerratum, g2-1, Zhao, Goerigk}). Similar studies exist for solid-state DFT as well \cite{Kurth, Vitos, Staroverov, Staroveroverratum, Heyd, Paier1, Tran, Grabowski, Ropo, Lany, Haas, Haaserratum, Csonka, Shang, Nazarov, Hautiererror}, but they are mostly limited to a small number of properties and/or compounds. In addition, their focus is often on understanding the differences between functionals, so they do not lead to quantitative and universally applicable error estimates with respect to experiment. A benchmark that is really comprehensive, should meet two criteria: the number of 
elements 
that 
is included in 
the test set 
should be sufficiently large, and the crystal structures in the set should be sufficiently diverse. This guarantees the transferability of the benchmark conclusions with respect to both the intrinsic and the numerical errors.

A natural choice to construct such an extensive test set emerges from the periodic table of elements. By taking the ground-state crystal structures of all elements \cite{Villars}, the two criteria for a comprehensive solid-state benchmark set are simultaneously fulfilled. All elements are included, thus trivially fulfilling the first requirement. In addition, the corresponding crystal structures range from simple hexagonal and cubic configurations to low-symmetry geometries, like orthorhombic and monoclinic cells. Of course, extrapolating the obtained insights to more complex materials, such as multicomponent compounds, requires some care. It is not impossible, however, as will be shown for example in Sec.~\ref{secstat}.

An additional advantage of using all elemental crystals follows from the periodic table's inherent ability to display trends and correlations. The systematic behavior of observable quantities along periods or groups is well known, but the deviations between DFT predictions and experimental values also appear to follow such trends (Sec.~\ref{secprop}). In particular, the largest errors are restricted to distinct regions. So apart from providing a complete test set, the classification of elements in the periodic table allows for an easy visualization and interpretation of the data. Furthermore, the elemental materials are among the best known and most studied materials on Earth. Experimental data collections are hence rather easy to find, and one can assume with sufficient confidence that the reported data are accurate.

By means of the ground-state elemental crystals, the present review offers an overview of the power and limitations of solid-state DFT calculations. Although this knowledge is often implicitly available in the computational community, most of it is scattered over two decades of literature. The current work therefore summarizes and quantifies this implicit knowledge into a practical protocol, that will allow any scientist -- experimental or theoretical -- to provide justifiable error estimates for many basic property predictions.  Intrinsic errors follow from a statistical analysis of the deviations between DFT predictions and experimental values for a given functional. Subsets of materials for which the deviations are particularly large are identified, and the reasons for this behavior are discussed. Numerical errors, on the other hand, express to what extent two independent DFT approaches produce identical predictions.
We will focus on the correspondence between two particular methods, APW+lo and PAW, by means of representative mainstream codes. For the PAW method, the use of different atomic potentials can have large effects, but by only considering the sets of potentials recommended by each code, we hope to establish a general idea of the PAW error.

Instead of performing an extensive level-of-theory study for various functionals, we will focus on one typical example within the generalized gradient approximation of DFT (GGA). For this, the PBE functional is chosen, because it is known to yield good results for solids of a wide range of elements and properties \cite{Csonka}. Moreover, its popularity \cite{DFTpoll} guarantees the comparibility and applicability of our results. Other GGA functionals are expected to display approximately the same behavior, except maybe for very specific material classes. Kurth \emph{et al.} \cite{Kurth} for instance have shown how four GGA functionals provide similar trends for both the equilibrium volume and the bulk modulus. Of course, the determination of the error estimates is not limited to GGAs. The presented methodology is also applicable to other functionals (such as LDA or hybrid functionals) or first-principles approches (such as Hartree-Fock, GW or RPA \cite{manybody}).

This review is organized as follows. Sec.~\ref{seccomp} describes the computational procedure for all properties under consideration and discusses the prerequisites for a sound comparison between theory and experiment. Within Sec.~\ref{secprop} the differences between DFT-GGA predictions and experimental values are assessed (intrinsic errors), whereas Sec.~\ref{seccode} focuses on the method and code dependence of the theoretically determined properties (numerical errors).

\section{Predicting experimental properties by means of DFT \label{seccomp}}

\subsection{Computational recipes \label{sechowprop}}

DFT computations for five distinct sets of materials properties will be discussed. They can be divided into energetic ($\Delta E_{coh}$) and elastic quantities ($V_0$, $B_0$, $B_1$, $C_{ij}$) \cite{Kittel}. Of course many more properties may be determined by means of DFT, but the quantities introduced here are directly available from straightforward total energy calculations. 

The \textbf{cohesive energy} or atomization energy $\Delta E_{coh}$ is a popular benchmark quantity \cite{PBE,g2-1,Kurth,Staroverov,Csonka,Vitos,Paier1,Tran}. Expressed as an energy difference per atom, it is defined as
\begin{equation}
 \Delta E_{coh} = - \left( E_0 - E_{at} \right) \label{eqEcoh}
\end{equation}
Here $E_0$ represents the energy per atom of the compound under investigation in its ground state, i.e. at 0\,K and without external stress. One can determine it by means of a standard pressure optimization procedure, or by fitting a few $E(V)$ data points to an empirical equation of state (EOS) and extracting the equilibrium energy analytically. In this work the latter option was chosen, using a common third-order Birch-Murnaghan relation \cite{Birch} :
\begin{multline}
 E(V) = E_0 + \frac{9 V_0 B_0}{16} \left\{ \left[ \left(\frac{V_0}{V}\right)^{2/3}-1 \right]^3 B_1 \right. \\ \left. + \left[ \left(\frac{V_0}{V}\right)^{2/3} - 1 \right]^2 \left[6-4\left( \frac{V_0}{V} \right)^{2/3} \right] \right\} \label{eqBM}
\end{multline}
$V_0$ represents the equilibrium volume, $B_0$ the bulk modulus and $B_1$ its pressure derivative. Other equations of state exist as well, but no significant difference with respect to ground-state properties is to be expected.

$E_{at}$ on the other hand is the energy of one isolated atom in its electronic ground state. Since many solid-state DFT codes only allow for the use of periodic boundary conditions, the isolated atom needs to be calculated in a periodic unit cell as well. In the present computations a free atom is placed in a big orthorhombic unit cell, such that every atom is surrounded by at least 15\,\AA \ of vacuum. In this way one can sufficiently suppress spurious interactions between periodic images ($< 1$\,meV). The orthorhombic symmetry is chosen over e.g. a simpler, cubic one, to avoid physically incorrect spherical states. After all, the use of a unit cell forces the electron density to assume the same symmetry as the lattice \cite{Philipsen1996}. This is in most cases only possible by means of partial occupation of the different electron orbitals, which is not physical. Lowering the crystal symmetry counteracts this phenomenon and should lead to strictly integer occupation numbers. However, some atoms end up 
with partially filled states, even when this approach is applied. In such cases, the occupation numbers have to be fixed manually before looking for the usual, self-consistent solution. In this work only experimental ground-state electron configurations are used: even when DFT predicts a different configuration to be more stable \cite{Baerends} (e.g. for W), the experimental occupation numbers are taken in order to guarantee a meaningful comparison to measurements. Only for spin-orbit coupled calculations for the Pb atom it is not possible to impose the experimental electronic state. The PBE ground state $^1$S$_0$ is therefore used, and not the experimental $^3$P$_0$ state.

The negative sign in Eq.~(\ref{eqEcoh}) causes positive cohesive energies to correspond to stable phases (with respect to atomic decohesion). The other sign convention, however, is commonly used as well.

One of the key physical properties of a given compound is its \textbf{volume}. In first-principles calculations the equilibrium volume per atom $V_0$ can be obtained easily. One either employs an optimization routine or fits some $E(V)$ points to an empirical equation of state. This is similar to the procedure used to determine $E_0$, and again the latter option is chosen in this work.

The \textbf{bulk modulus} is closely related to the $E(V)$ behavior as well. It is proportional to the curvature of the equation of state at the equilibrium volume:
\begin{equation}
 B_0 = - \left. V \frac{\partial P}{\partial V} \right|_{V=V_0} = V \left. \frac{\partial^2 E}{\partial V^2} \right|_{V=V_0} \label{eqB0}
\end{equation}
It represents the resistance of the unloaded material to volume change, and hence to uniform pressure. Because it is linked to the curvature of the $E(V)$ relation, $B_0$ is a numerically sensitive quantity. A small deviation at a few data points is already able to change its value noticeably, especially when the bulk modulus is small (shallow EOS). This is increasingly so when only a narrow volume range is inspected. 

$B_1$ stands for the \textbf{derivative of the bulk modulus} with respect to pressure, evaluated at the equilibrium volume:
\begin{equation}
 B_1 = \left. \frac{\partial B}{\partial P} \right|_{V=V_0} = \left. \frac{\partial}{\partial P} \left( V \frac{\partial^2 E}{\partial V^2} \right) \right|_{V=V_0}
\end{equation}
It is a third-order derivative of the energy and hence describes effects that are one order higher even than the bulk modulus. It is related to the volume-dependence of the $E(V)$ curvature. $B_1$ is therefore the most sensitive elastic quantity discussed in this study. Again, both the bulk modulus and its pressure derivative are obtained from fitting an EOS to calculated $E(V)$ data points.

The mechanical behavior of a crystal cannot be described solely by means of the bulk modulus. When anisotropic deformations are applied, other \textbf{elastic constants} come into play as well. The full set of these constants makes up the stiffness matrix $\vt{C}$. It represents a tensor of rank 2 and relates (small) cell strains to the corresponding stresses via Hooke's law $\vt{\sigma} = \vt{C} \cdot \vt{\epsilon}$. $\vt{C}$ is a symmetric $6\times6$ matrix, containing 21 independent constants at the most. In the case of hexagonal crystals five distinct values remain ($C_{11}$, $C_{12}$, $C_{33}$, $C_{13}$, and $C_{44}$), while for cubic compounds there are only three ($C_{11}$, $C_{12}$, and $C_{44}$). The $C_{ij}$ parameters can also be translated into more general elastic moduli, such as Young's modulus $E$, the shear modulus $G$ and Poisson's ratio $\nu$. Even the bulk modulus can be obtained from a simple combination of the $C_{ij}$. In addition, the elastic constants are known to relate to structural 
stability and various other important physical properties \cite{Kim2009, Miao}.

Several methods are available to obtain the elastic constants from first principles, either by relating energy and strain \cite{LePage2001} or stress and strain \cite{LePage, Shang2007}. In most cases a stress-based procedure is preferred, because it is inherently faster. However, it requires an \emph{ab initio} code that can determine the stress tensor. In a first step the cell pressure components are then extracted for a minimal set of deformed geometries. Together with the corresponding strains, this results in a system of linear equations. Solving that system yields the required elastic constants. When it is important to obtain an accurate value of $C_{ij}$, one should construct an overdetermined system, by applying the same strain sets at different magnitudes. The elastic constants can then be retrieved by using a least-squares method.

\subsection{Comparing theory and experiment \label{seccorrections}}

When a DFT prediction is compared to a number from experiment, the corresponding ambient conditions should be as identical as possible. This means in the first place that the experimental result should refer to 0\,K. Moreover, the measurement should be corrected for zero-point vibrational effects, which are not present in standard DFT calculations. The following paragraphs discuss how to extrapolate the experimental values to absolute zero and correct them for zero-point vibrations.

For the \textbf{cohesive energy} it takes little effort to match up theory and experiment consistently. Experimental data at low temperatures are in most cases available. Only the zero-point energy $\zeta$ hinders a direct comparison between 0\,K and experiment. From quantum mechanics this quantity is known to be $\frac{3}{2} \hbar \langle \omega \rangle$, with $\langle \omega \rangle$ the average phonon frequency. The latter can be estimated from Debye theory, where it is proportional to the maximum vibrational frequency, and hence to the Debye temperature $\Theta_D$. The zero-point energy correction becomes \cite{Alchagirov}
\begin{equation}
 \zeta = \frac{9}{8} k_B \Theta_D \label{zeta}
\end{equation}
Theoretical cohesive energies can only be compared to experiment if this contribution is added to the experimental values (added, due to the chosen sign convention in Eq.~(\ref{eqEcoh})). 

When no experimental value is available, $\Theta_D$ can be estimated. Here the Debye-Gr\"uneisen approximation \cite{Moruzzi}
\begin{equation}
 \Theta_D = 0.617 \frac{\hbar}{k_B} \left(6 \pi^2 \right)^{1/3} V_0^{1/6} \left( \frac{B_0}{M} \right)^{1/2} \label{eqThetaD}
\end{equation}
will be used. Both $V_0$ and the mass $M$ are expressed per particle, corresponding to a single atom for most materials. For dimeric crystals, however, the diatomic molecule is chosen as a unit of repetition. The regular, room temperature experimental values for $B_0$ and $V_0$ are filled in, except when the difference with low temperature results (see further) is significant. This is the case for Cl, Br, and I.  

Thermal \textbf{volume} corrections consist of two parts. Assuming to have a room temperature measurement at one's disposal, the first step consists in accounting for thermal expansion from absolute zero to ambient temperature:
\begin{equation}
 \frac{\Delta V^{(1)}}{V} = \int_{0}^{T_{rt}} \alpha_V (T) \, \mathrm{d}T \label{eqdV1a}
\end{equation}
$\alpha_V(T)$ represents the temperature-dependent volume expansion coefficient. It is zero at 0\,K and $\alpha_{V,rt}$ at room temperature ($T_{rt}$). Since $\Delta V^{(1)}$ constitutes only a small correction with respect to the total volume $V$, Eq.~(\ref{eqdV1a}) will be approximated here as
\begin{equation}
 \frac{\Delta V^{(1)}}{V} \approx \int_0^{T_{rt}} \alpha_{V,rt} \, \frac{T}{T_{rt}} \, \mathrm{d}T = \frac{\alpha_{V,rt} T_{rt}}{2} \label{eqdV1b}
\end{equation}
In a limited number of cases the experimental expansion coefficient is not known. It can then be estimated from an empirical correlation to the `moleculization' energy \cite{Tsuru}:
\begin{equation}
 \alpha_{V,rt} = 3 \times \frac{48.14 \cdot 10^{-6} \, \mathrm{eV/K/atom}}{\Delta E_{mol}} \label{eqalpha}
\end{equation}
$\Delta E_{mol}$ is defined as the energy difference per atom between the crystalline material and its gaslike molecules. For elements with an atomic gas phase, it reduces to the atomization energy (cohesive energy). In the absence of experimental data on both $\alpha_{V,rt}$ and $\Delta E_{mol}$, Eq.~(\ref{eqalpha}) is completed with DFT values.

A second modification is again due to zero-point effects. Because of the volume-dependence of the zero-point energy $\zeta$, the equilibrium volume is shifted slightly. According to Alchagirov \emph{et al.} \cite{Alchagirov, Haas}, this small difference per atom amounts to
\begin{equation}
 \Delta V^{(2)} = \frac{(B_1-1) \zeta}{2 B_0} = \frac{9}{16} (B_1 - 1) \frac{k_B \Theta_D}{B_0} \label{dV2}
\end{equation}
Dacorogna and Cohen \cite{Dacorogna} propose an alternative definition of the zero-point volume shift. They obtain a similar formula, but with $B_1$ instead of $B_1-1$ in Eq.~(\ref{dV2}). However, the mathematical expression is preceded by some significant simplifications. When calculating zero-point effects it is therefore advisable to use Eq.~(\ref{dV2}) instead, especially when $B_1$ is small.

For the \textbf{bulk modulus} thermal effects should be taken into account as well.  A first contribution originates in the thermal expansion of the material. Similar to $\Delta V^{(1)}$, a correction $\Delta B^{(1)}$ can be determined too. Roughly approximating the relevant behavior, one can write \cite{Gaudoin}
\begin{equation}
 \Delta B^{(1)} = B_1 \cdot P \left( \Delta V^{(1)} \right) = - B_0 B_1 \frac{\Delta V^{(1)}}{V_0} \label{dB1}
\end{equation}

\begin{table*}[t]
 \caption{Ground-state crystal structures for all elements up to radon. Both the space group number and the Pearson notation are given (with hR$x$ standing for $x$ atoms in the \emph{hexagonal} setting of the rhombohedral unit cell) \label{tabstruct}}
 \vspace{1ex}
 \includegraphics[width=\textwidth]{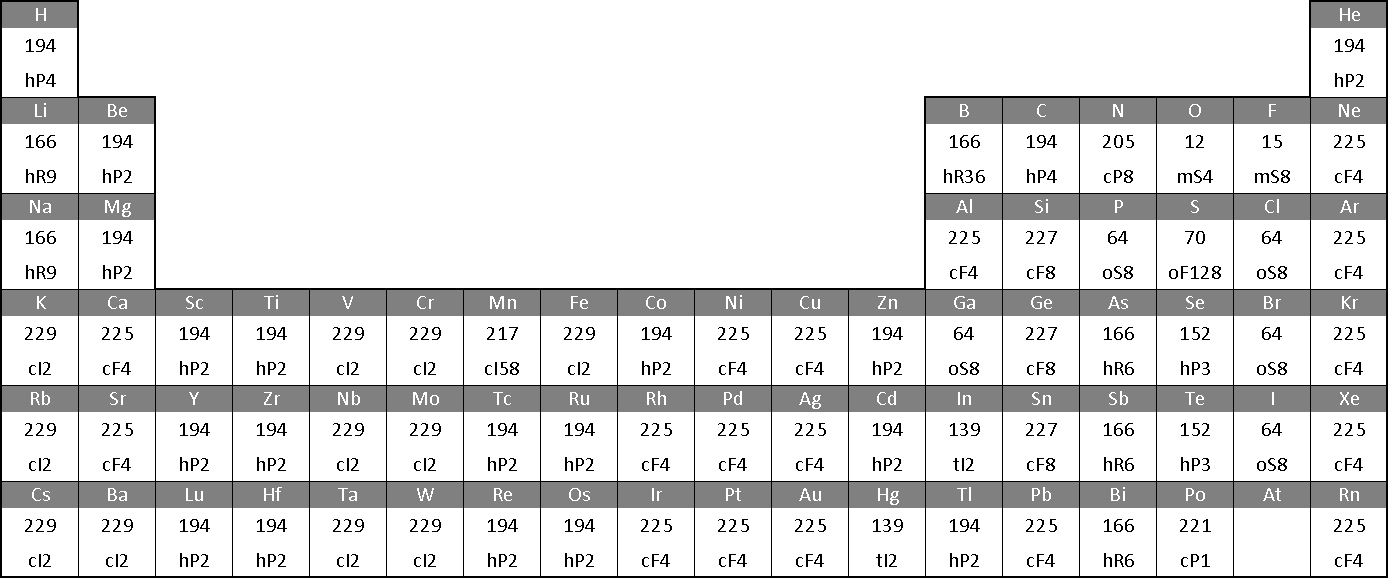}
\end{table*}

On the other hand, the effect of zero-point vibrations on the bulk modulus boils down to \cite{Alchagirov}
\begin{multline}
 \frac{\Delta B^{(2)}}{B_0} = - \frac{\Delta V^{(2)}}{V_0} \left[ \frac{1}{2} \left(B_1 - 1 \right) \right. \\ \left. + \frac{2}{B_1-1} \left( \frac{2}{9} - \frac{1}{3} B_1 - \frac{1}{2} B_0 B_2 \right) \right] \label{dB2}
\end{multline}
$B_2$ stands for the second-order derivative of the bulk modulus with respect to pressure. It is a highly sensitive parameter and very difficult to extract from a few $E(V)$ data points. In addition, $B_2$ is not included in Eq.~(\ref{eqBM}) and a higher-order Birch-Murnaghan fit should be applied. Instead, the present work will use the intrinsic Birch-Murnaghan value:
\begin{align}
 \left(B_0 B_2 \right)^{BM} &= B_0 \left. \frac{\partial^2}{\partial P^2} \left( V \frac{\partial^2 E^{BM}}{\partial V^2} \right) \right|_{V=V_0} \nonumber \\
 &= - \frac{143}{9} + 7 B_1 - B_1^2
\end{align}
There are other possibilities as well \cite{Alchagirov}, ranging from a different equation of state to an accurate numerical determination of $B_2$. In order to establish the small correction $\Delta B^{(2)}$, however, this more consistent approach suffices. 

Since it is already hard to accurately measure a high-order parameter like $B_1$ or to determine it from first principles, zero temperature modifications will often be negligible compared to experimental or computational errors. $B_1$ is therefore not adjusted to incorporate thermal expansion or zero-point effects. 

No thermal corrections are applied to the elastic constants $C_{ij}$ as well. One can however imagine a modification similar to that of Dacorogna and Cohen \cite{Dacorogna} for the bulk modulus:
\begin{align}
 \Delta C_{ij}^{(m)} &= \frac{\partial C_{ij}}{\partial P} \cdot P \left( \Delta V^{(m)} \right) \nonumber \\
 &= - B_0 \frac{\partial C_{ij}}{\partial P} \frac{\Delta V^{(m)}}{V_0}
\end{align}
with $m=1$ to account for thermal expansion and $m=2$ for zero-point effects. Unfortunately experimental data about the pressure derivative of the elastic constants are scarce.

\section{Intrinsic errors \label{secprop}}

\subsection{Test set preparation \label{secpropcomp}}

In order to establish statistically justified intrinsic error estimates, the ground-state elemental crystals at 0\,K will be used as a benchmark set. Pettifor \cite{Pettifor} lists these crystal structures, based on an overview by Villars and Daams \cite{Villars}. However, in some cases literature suggests another phase to be even more stable at low temperatures. In order to ensure the use of 0\,K cell geometries as much as possible, such an alternate structure is taken for boron \cite{Masago}, nitrogen \cite{Scott}, oxygen \cite{oxygen}, and sulfur \cite{David}. Tab.~\ref{tabstruct} presents an overview of all structures used in the current test set.

Using a 0\,K benchmark set entails two distinct advantages. On the one hand some elements only crystallize just above absolute zero. Collecting both 300\,K and 0\,K compounds in one set might then seem a bit inconsistent. On the other hand this approach facilitates the extrapolation from the experimental temperature to 0\,K, as there are no phase transformations along the way.

All structures are considered in their stress-free ground state. This means that, when the space group allows some freedom in the internal positions, an optimization with respect to the total energy is necessary. This optimization procedure calls for a fast and well-accepted DFT algorithm. The projector augmented wave method \cite{Bloechl, potpaw} (PAW) as implemented in VASP \cite{vasp, vasp2} (version 5.2.2) fulfills both criteria. The elemental crystal structures have therefore been relaxed by means of this code, using the recommended PAW atomic potentials listed in the manual \cite{VASPmanual}. A force convergence criterion of 0.01\,eV/\AA \ was set. All calculations have been performed using the tetrahedron method with Bl\"ochl corrections \cite{Bloechl2}, while the reciprocal space was sampled by means of a Monkhorst-Pack grid \cite{MonkhorstPack}. Further computational details for the calculations are given in the Supplementary Material \cite{suppl}.

The equilibrium structure has been obtained in two stages. For the determination of the equilibrium volume a uniformly spaced 13-point EOS (up to $V_0 \pm 6$\,\%) has been calculated and fitted to a least-squares third order Birch-Murnaghan relation (see Sec.~\ref{sechowprop}). Only for a number of shallow $E(V)$ curves --- in particular for H, N, S, the halogens, and the noble gases --- an increased volume range turned out to be necessary. For each of the 13 crystal volumes, the atomic positions and the cell shape have been individually optimized. In a second step, the crystal has been reinitialized at the fitted $V_0$ and has then been optimized again.

These optimized crystal structures form the definitive test set (submitted to the COD \cite{cod} and ICSD \cite{icsd} crystallographic databases). For each of them, most of the properties discussed in Sec.~\ref{sechowprop} have been determined in order to quantify the difference between PBE and experimental values (Sec.~\ref{secpred}). The DFT part of the comparison has been performed by means of VASP, using the settings mentioned earlier. They allow to converge all energy differences up to a few meV per atom at the most. For O and Cr (antiferromagnetic), Mn (ferrimagnetic), Fe, Co, and Ni (ferromagnetic), spin polarization has been taken into account, while for the heaviest elements (as from Lu) spin-orbit contributions have been incorporated. At that point relativistic effects beyond the scalar-relativistic approach become important, as will be shown later (Tab.~\ref{tabrel}).

The analysis in Sec.~\ref{secpred} will not show the raw calculated data, but will rather elaborate on the deviation between theory and experiment. The first-principles results and the thermally corrected experimental numbers \cite{Kittel,Greenwood,Tohei,Villars,Bolz,Meyer,Batchelder,Powell,Peterson,Losee,Sears,Swenson,graphite,Scott,Endoh,Duesing,Keeler,Fujihisa,Skalyo,Price,Lurie,Knittle,Batsanov,Guinan,Menoni,Beister,Kenichi,Libp,Shang,Hector,Buchenau,CRC,Nelson,Keyes,Wallis,Pawar,Anderson1983} have been included in the Supplementary Material \cite{suppl}. A tabulation of calculated and experimental values for the elastic constants $C_{ij}$ was published before by Shang \emph{et al.} \cite{Shang} for most of the present benchmark set. In Sec.~\ref{secpred} their data are used. Only for the experimental numbers of Ba we found more realistic results elsewhere \cite{Hector, Buchenau}. Since the authors only considered bcc, fcc and hcp structures, this implies that for Li and Na a different geometry was 
applied than in the rest of this work (bcc instead of hR9). Moreover their results are based on a PW91 functional \cite{PW91a,PW91b}, rather than the PBE approximation employed in the rest of this work. Although these GGA approaches yield different results in a few situations \cite{Mattsson}, they are in most cases very similar and for the elastic constants no significant deviation is expected.

\subsection{Statistical analysis \label{secstat}}

\subsubsection{Linear regression}

Benchmark studies usually analyze the difference between DFT and experiment statistically. The most common characteristics investigated are the mean error (signed) and the mean absolute error (unsigned). However, this approach implicitly assumes that the offset between DFT predictions and experimental results is the same for large and small values. For strictly positive quantities a relative shift seems more reasonable. The present analysis therefore explicitly treats relative deviations, in addition to the remaining scatter on that trend. This is done by means of a linear regression between DFT data and experimental results.

\begin{figure}[t]
 \includegraphics[width=0.68\columnwidth]{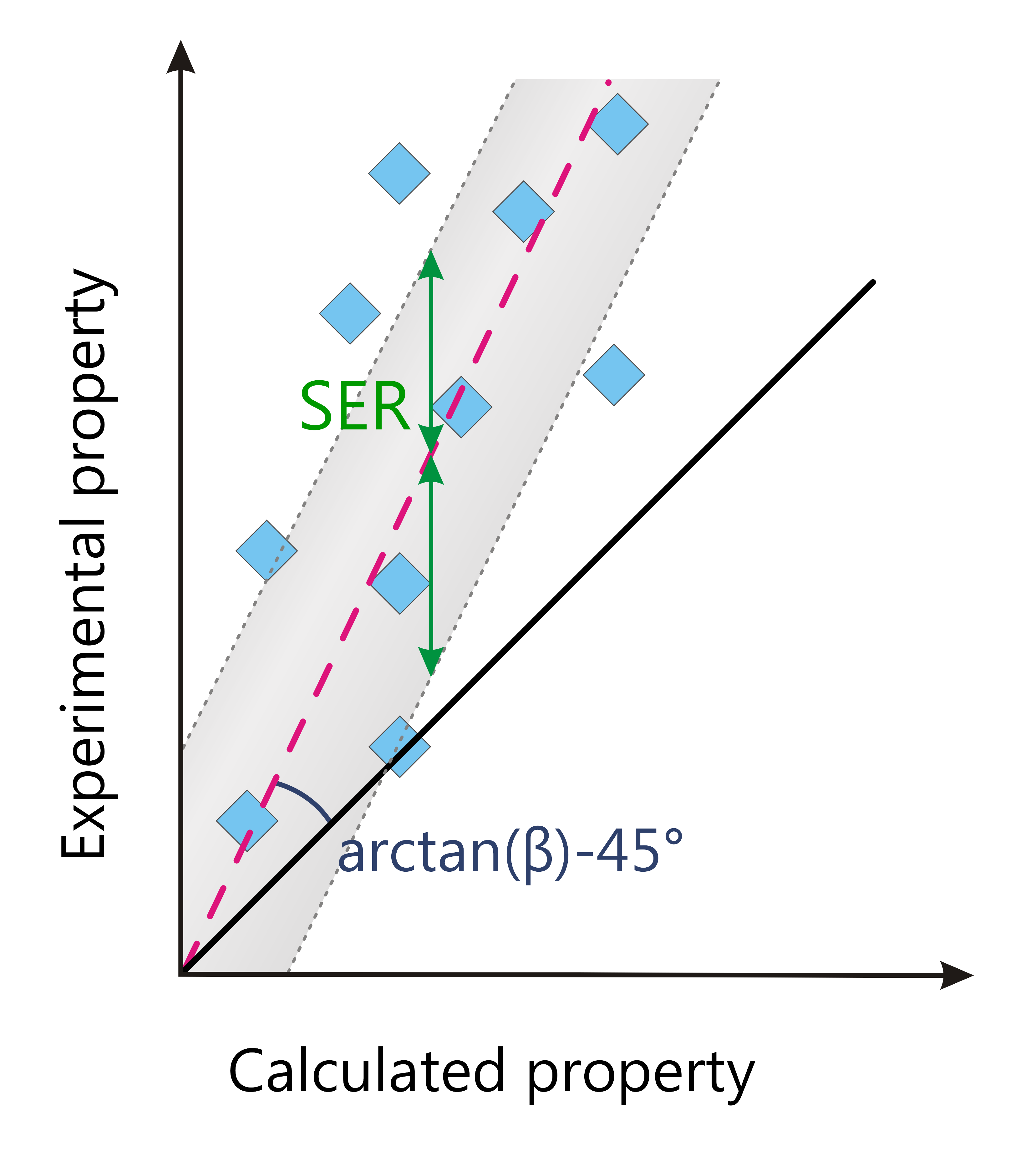}
 \caption{(Color online) The intrinsic error between experiment and DFT can be decomposed in a systematic deviation and a residual error bar. Systematic differences are quantified by comparing the linear regression line $X = \beta T$ (red, dashed) to the bisector $X = T$ (black, full). The residual error bar (green arrows) is defined as the standard deviation of the regression errors (SER) \label{figerrorbar}}
\end{figure}

The linear regression is performed by means of a least-squares fit, from which we obtain the slope as well as the scatter with respect to the regression line \cite{DeGroot}. The model hence presumes that a perfect correlation between experiment $X$ and theory $T$ exists, distorted by a random error $\epsilon$ centered around a zero mean: $X=\beta T + \epsilon$. If the exact exchange-correlation functional were known, it would approximately lead to $\epsilon=0$ and $\beta=1$. In practice, comparing the least-squares estimate of $\beta$ to 1 offers a good measure of any systematic deviations, while the standard deviation of $\epsilon$ (also denoted as standard error of the regression or SER) expresses the residual error bar (see Fig.~\ref{figerrorbar}).

Although the experimental community commonly employs the nomenclature `systematic errors' and `non-systematic errors', we will avoid using the second term. Non-systematic errors suggest a certain degree of randomness, which is not present in DFT. When one performs the same experiment several times, the results are spread around a mean value. In DFT, repeating the same calculation always yields identical results. In contrast to experimental error analysis, the spread of deviations in DFT results becomes apparent only when predictions for many different compounds are compared to experiment, i.e. when a benchmark set is used. The DFT scatter is then caused by some (subsets of) crystals being described better by the functional than others. Intrinsic error bars are hence fundamentally different from experimental error bars as well. The systematicness of DFT also appears when studying results within a certain family of materials: similar systems behave almost identically, proving the DFT spread
not to 
behave randomly at all. An excellent example is found in literature, where the correspondence with experiment 
displays much less scatter when chemically similar compounds are involved \cite{Hautiererror}.

As an additional side note, one should be aware that experimental error bars have not been taken into account in any way. In fact, the statistical model should be $X + \eta =\beta T + \epsilon$, where $\eta$ represents an additional (but uncorrelated) zero-mean perturbation. This does not only affect the comparison between individual DFT and experimental results, but also influences the values of the intrinsic systematic deviations and residual error bars that are presented in this work. By considering a test set that is sufficiently large, such as in the present study, one may hope that these effects level out. In addition, the elemental crystals belong to the most intensively studied materials, such that the experimental errors are much smaller than for regular compounds. Without full knowledge of the experimental errors, however, we can only state that the SER provides an upper limit for the real PBE spread $\sigma_\epsilon$. A possible solution consists in comparing DFT values to results from 
highly accurate many-body techniques instead. Such high-precision data are not available for many of the materials considered here, however, and to calculate them ourselves 
exceeds the scope of the current review.

\subsubsection{Eliminating outliers}

A full statistical analysis of all elemental data cannot be performed straightforwardly. Some subsets of elements strongly distort the agreement between DFT and experiment. More meaningful error estimates are obtained when the most striking outliers are removed from the data set. Since the deviating behavior is often caused by a bad description of some underlying physical mechanism, most of them will be grouped in subsets of similar compounds. Instead of removing one outlier at a time from the data set, we choose only to remove entire structure types at once. This way, individual materials that behave well for the wrong reasons, are excluded as well, avoiding bias towards smaller errors. 

\begin{figure}[t]
 \includegraphics[width=\columnwidth]{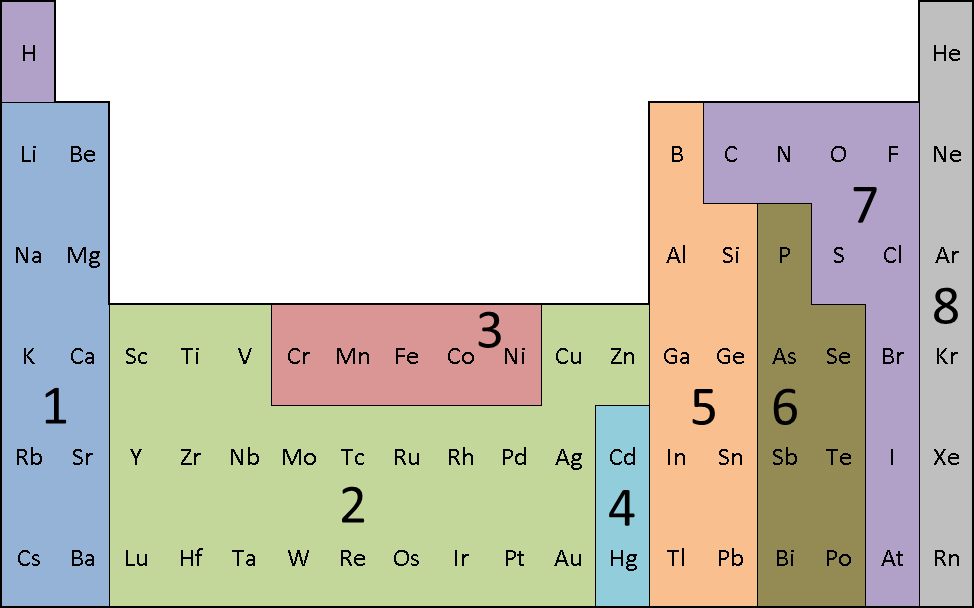}
 \caption{(Color online) Decomposition of the periodic table into smaller subsets of elements, based on common physical properties of the corresponding ground-state crystals: (1) alkali and alkaline earth metals, (2) nonmagnetic transition metals, (3) magnetic materials, (4) correlation-dominated materials, (5) high-coordination $p$ block compounds, (6) low-coordination $p$ block compounds, (7) molecular crystals, and (8) noble gases. Subsets 7 and 8 correspond to materials where dispersion interactions are essential \label{figgroups}}
\end{figure}

A decomposition of the test set into eight subsets is proposed, based on some common physical properties of the corresponding elemental crystals (Fig.~\ref{figgroups}). They are: (1) alkali and alkaline earth metals, (2) nonmagnetic transition metals, (3) magnetic materials, (4) correlation-dominated materials, (5) high-coordination $p$ block compounds, (6) low-coordination $p$ block compounds, (7) molecular crystals, and (8) noble gases. Obviously for some boundary elements, the most appropriate subset can be matter for discussion, but the classification in Fig.~\ref{figgroups} explains most trends for the intrinsic errors in a satisfactory manner (see further).

These eight subsets of elemental materials are representative for more complex (multicomponent) crystals as well. They provide prototype systems for particular bond types and physical phenomena, such as London dispersion (subsets 7 and 8), magnetism (subset 3) and electronic correlation (subset 4). Observations of DFT performance for these eight subsets will therefore carry over to multicomponent compounds.

In order to eliminate deviating subsets in an objective way, the following procedure has been used. All subsets from Fig.~\ref{figgroups} that have half or more of their elements differing significantly from the dominating trend, have been excluded. A two-sided p-value of 10\,\% is maintained. In other words, a data point is considered to deviate substantially when the (signed) relative residual error (expressing the difference between the DFT regression value and the experimental one) belongs to the outer 10\,\% of a normal distribution. This approach has been repeated in an iterative way: after the elimination of each subset the significance criterion has been reestablished, until no deviating subsets remained. For solids belonging to an excluded category, PBE is not expected to provide reliable property predictions.

This selection criterion has been visualized in Tabs.~\ref{taberrors1} to \ref{taberrors2}. For each elemental crystal the relative residual error $|x_{exp}-x_{reg}|/x_{exp}$ is shown. Large numbers suggest a significant deviation from the regression line and hence allow to recognize outliers. Because these differences are displayed in the shape of the periodic table, they allow for easy identification of deviating subsets\footnote{The graphical representations in this work employ the conventional, medium-form periodic table. Hydrogen is kept in group IA (contrary to the vivid discussion in e.g.\ Ref.~\onlinecite{Scerri2010}) and lutetium in group IIIB \cite{Scerri2009}. This increases the intuitive character of the results. Insights should be conveyed at a glance and many researchers are most familiar with the standard format of the periodic table. For the same reason the two-dimensional representation is preferred over some 1D alternatives \cite{Villars2001}.}.
\nocite{Scerri2010,Scerri2009,Villars2001} A color code has been added to improve intuition, with the darkest shades corresponding to the largest deviations. The deviations with respect to the elastic constants represent the mean absolute errors over $C_{11}$, $C_{12}$, $C_{33}$, $C_{13}$, and $C_{44}$. 

\begin{table*}[p]
 \caption{(Color online) Relative residual errors (of the VASP-PBE regression results with respect to the thermally corrected experimental values) for $\Delta E_{coh}$ \cite{Kittel,Greenwood,Tohei} (green), $V_0$ \cite{Villars,Bolz,Meyer,Batchelder,Powell,Peterson,Losee,Sears,Swenson} (red) and $B_0$ \cite{Kittel, graphite, Scott, Endoh, Duesing, Keeler, Fujihisa, Skalyo, Price, Lurie} (blue) of the elemental crystals. The darkest shades correspond to the largest errors \label{taberrors1}}
 \vspace{1ex}
 \includegraphics[width=\textwidth]{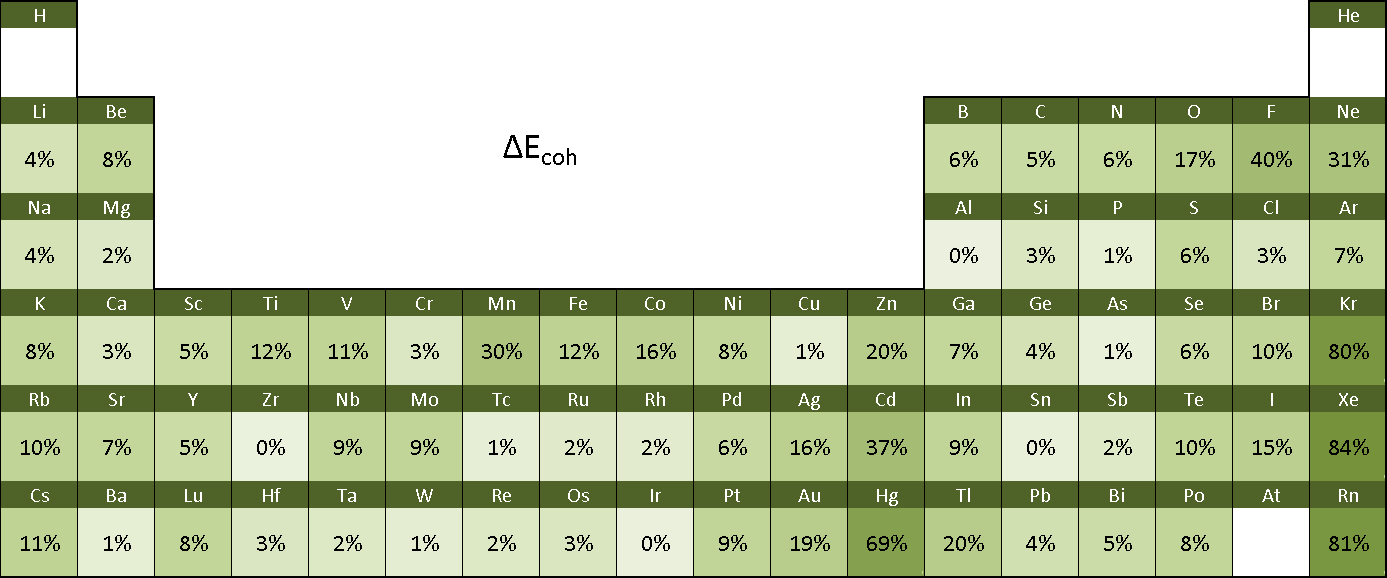} \\ \vspace{0.5cm}
 \includegraphics[width=\textwidth]{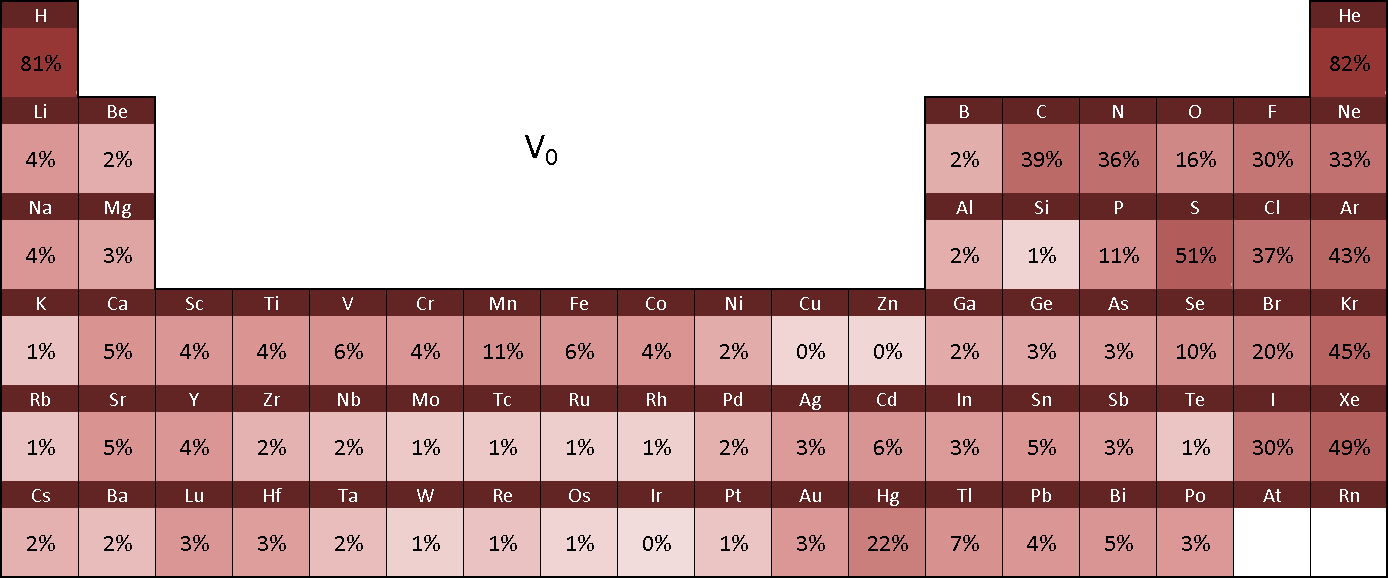} \\ \vspace{0.5cm}
 \includegraphics[width=\textwidth]{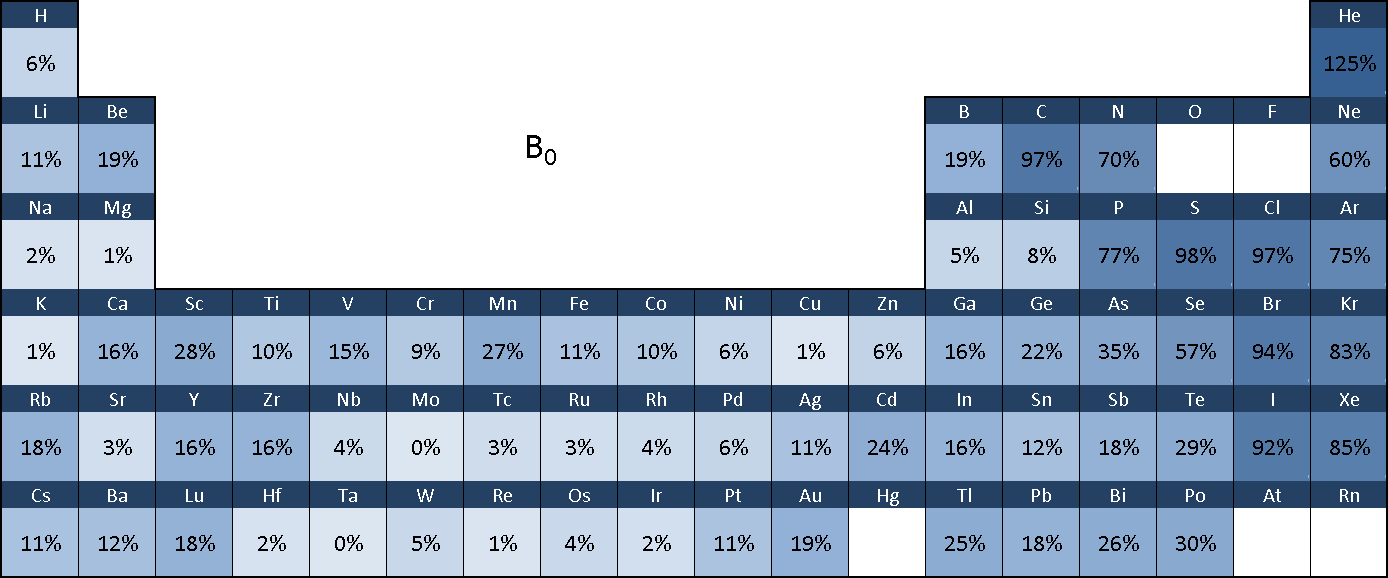}
\end{table*}

\begin{table*}[t]
  \caption{(Color online) Relative residual errors (of the VASP-PBE / -PW91 regression results with respect to the thermally corrected experimental values) for $B_1$ \cite{Knittle, Batsanov, Guinan, Menoni, Beister, Kenichi, Libp} (purple) and $C_{ij}$ \cite{Shang,Hector,Buchenau} (PW91, cyan) of the elemental crystals. The darkest shades correspond to the largest errors \label{taberrors2}}
 \vspace{1ex}
 \includegraphics[width=\textwidth]{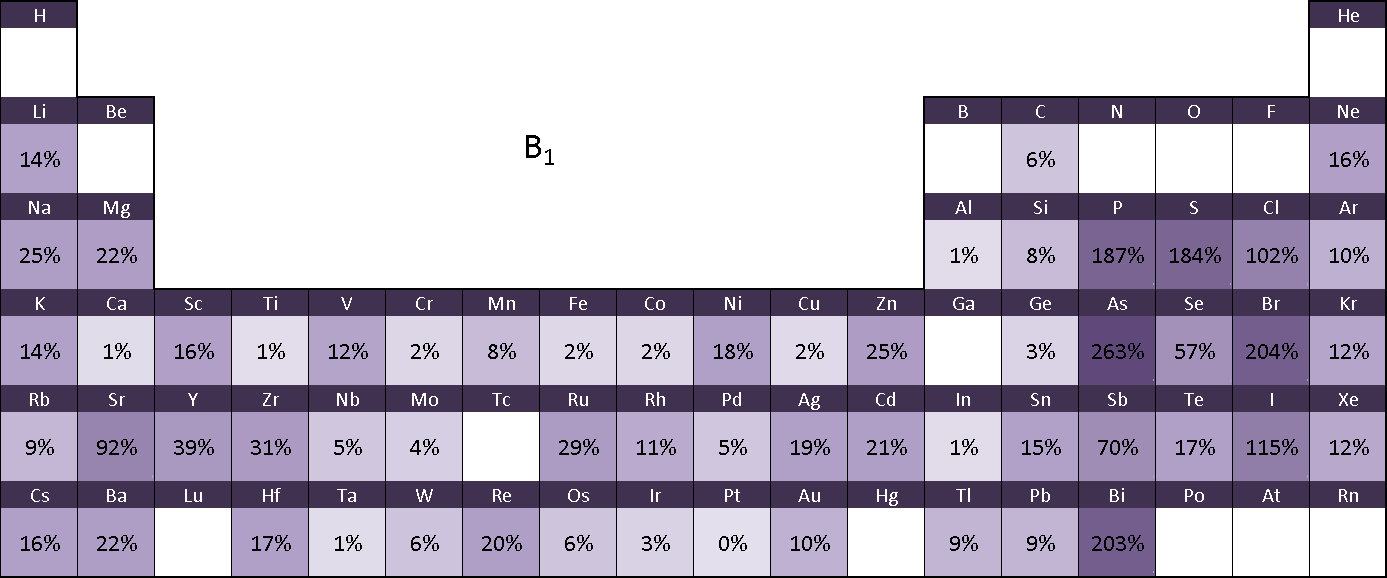} \\ \vspace{0.5cm}
 \includegraphics[width=\textwidth]{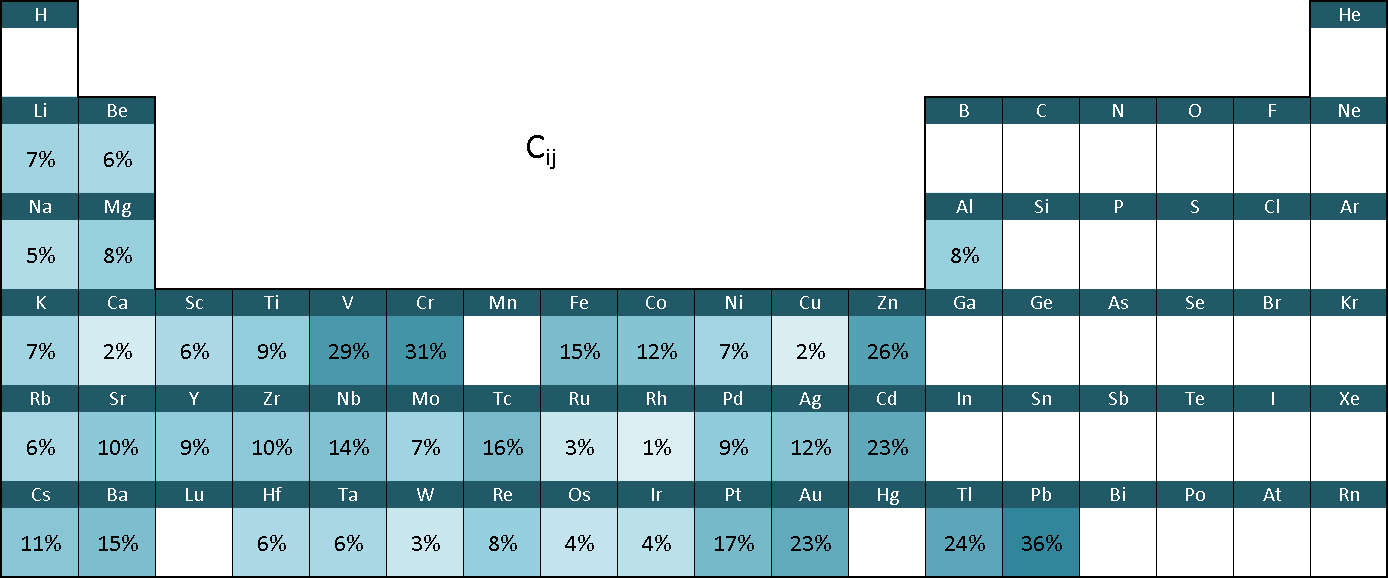}
\end{table*}

Another way to visualize the assessment procedure is presented in Fig.~\ref{figreg1}. In order to get a better view on systematic deviations, both the linear regression line (dashed red) and the first quadrant bisector (full black, representing a perfect match between theory and experiment) have been added for all accepted elements (see Tab.~\ref{tabaverror}). As an additional quality indicator, the Pearson product-moment correlation coefficient $r$ has been included for each property \cite{Rodgers}, with $r=1$ indicating a perfect positive correlation. Data points corresponding to omitted subsets, on the other hand, have been represented by an open symbol.

\subsubsection{Predicting experiment \label{secpred}}

\begin{figure*}[p]
 \subfigure[]{\includegraphics[height=0.37\textwidth]{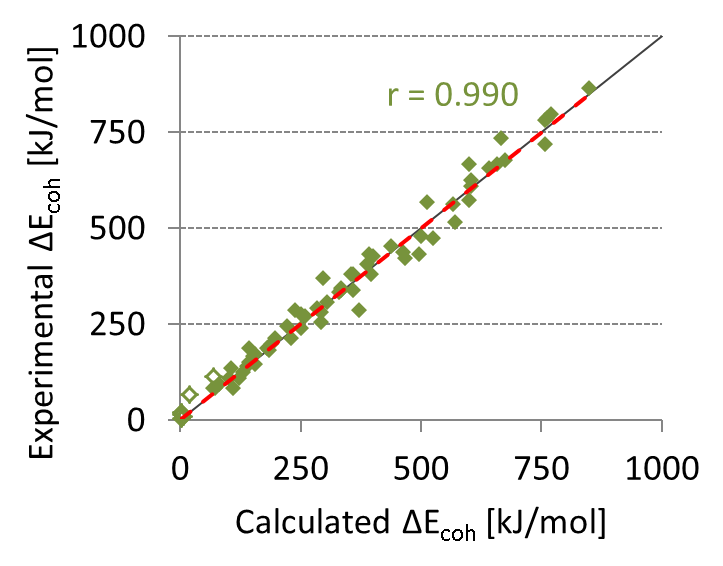} \label{figregEcoh}} \hfill \subfigure[]{\includegraphics[height=0.37\textwidth]{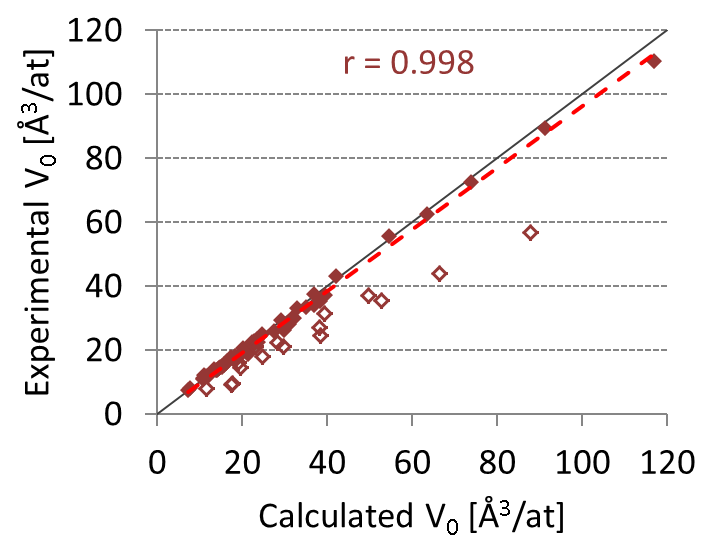} \label{figregV}} \\
 \subfigure[]{\includegraphics[height=0.37\textwidth]{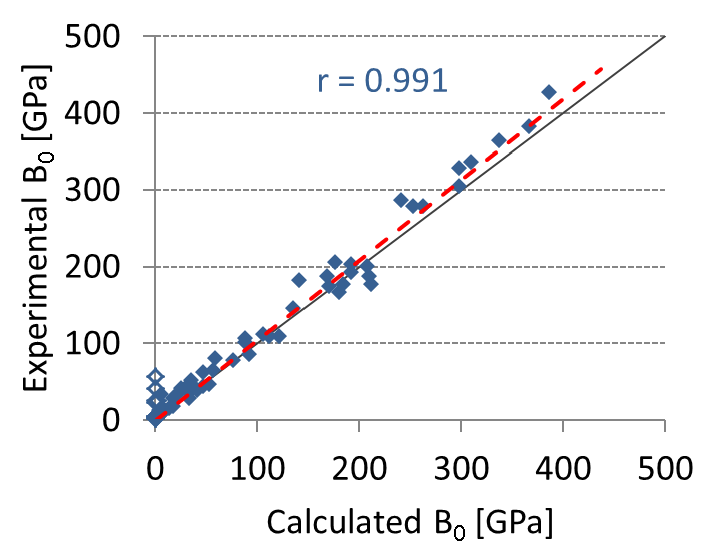} \label{figregB}} \hfill \subfigure[]{\includegraphics[height=0.37\textwidth]{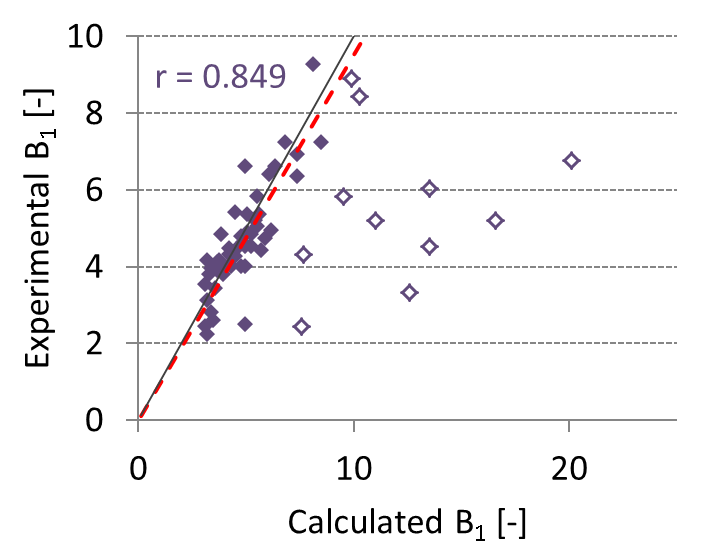} \label{figregBP}} \\
 \subfigure[]{\includegraphics[height=0.37\textwidth]{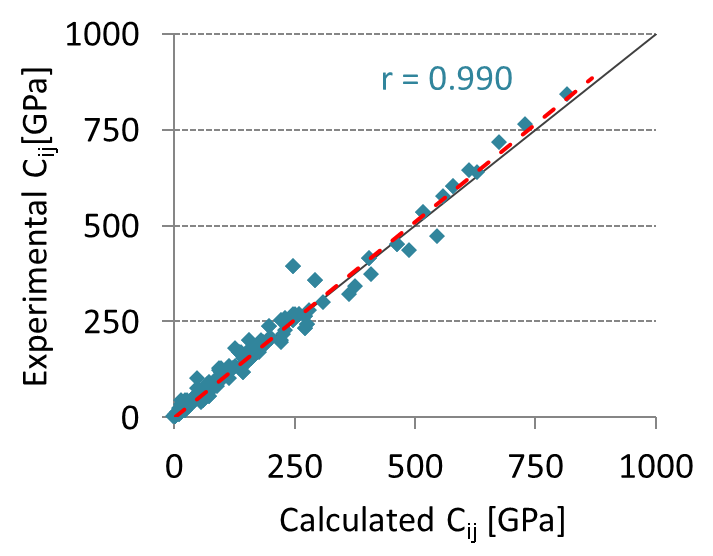} \label{figreg2}} 
 \caption{(Color online) Linear regression (dashed red) between the (thermally corrected) experimental and theoretical results (VASP-PBE / -PW91, see text) for the cohesive energy \cite{Kittel,Tohei}, equilibrium volume \cite{Villars,Bolz,Powell,Batchelder,Peterson,Losee,Sears,Swenson,Meyer}, bulk modulus \cite{Kittel, Duesing, graphite, Scott, Endoh, Keeler, Skalyo, Lurie, Fujihisa, Price}, pressure derivative of the bulk modulus \cite{Knittle, Batsanov, Guinan, Beister, Menoni, Kenichi, Libp}, and elastic constants \cite{Shang, Hector, Buchenau}. The full, black line stands for $x_{exp}=x_{th}$. $r$~represents the Pearson product-moment correlation coefficient for all elements included in the regression (filled symbols). The criterion for excluding certain elements from the fit (open symbols) is discussed in the text \label{figreg1}}
\end{figure*}

Using all crystals that survived this selection procedure (filled symbols in Fig.~\ref{figreg1}), a least-squares linear regression can now be computed for all properties from Sec.~\ref{sechowprop}. Tab.~\ref{tabaverror} summarizes the resulting intrinsic errors in terms of relative systematic deviations ($1-\beta$) and residual error bars (SER). Systematic errors are mentioned for PBE with respect to experiment, so a positive number implies PBE to overestimate that property. Each percentage is expressed relative to the PBE result, allowing for a straightforward calculation of the regression value ($x_{reg} = x_{th} - (1-\beta) x_{th}$). Between brackets the significance level of $\beta \not= 1$ is mentioned. It represents the two-sided p-value when a null hypothesis of $\beta=1$ is assumed: if there really were no deviation at all, a small p-value would indicate that finding an even more extreme result would be highly unlikely. For the residual error bars a 95\,\% confidence interval is 
given. The last column lists the subsets of elements that were excluded from the regression analysis by the selection procedure described before. For this the naming convention from Fig.~\ref{figgroups} has been used. 

\begin{table}[t]
 \caption{Systematic deviations $1-\beta$ and intrinsic error bars (SER) for the VASP-PBE/-PW91 (see text) properties presented in Tabs.~\ref{taberrors1}-\ref{taberrors2}, compared with experiment. The significance of the systematic deviation from $x_{exp}=x_{th}$ is indicated between brackets, by means of the two-sided $p$-value (low $p$-values for high significance). For the standard error of the regression a 95\,\% confidence interval is given in terms of the upper and lower limit (superscript and subscript). Subsets containing a lot of outliers have been excluded from the data set by means of the procedure mentioned in the text (notation from Fig.~\ref{figgroups}) \label{tabaverror}} \vspace{0.3cm}
 \begin{tabular}{l|c|c|r}
  \hline \hline & $1-\beta$ & SER & excl. \\
  \hline $\Delta E_{coh}$ [kJ/mol] & $-0.0\,\%$ (0.99) & 30 $_{-4}^{+7}$ & 4, 8 \\
  $V_0$ [\AA$^3$/atom] & $+3.6\,\%$ ($10^{-10}$) & 1.1 $_{-0.2}^{+0.2}$ & 4, 7, 8 \\
  $B_0$ [GPa] & $-4.9\,\%$ ($10^{-3}$) & 15 $_{-2}^{+4}$ & 7, 8 \\
  $B_1$ [--] & $+4.8\,\%$ (0.03) & 0.7 $_{-0.1}^{+0.2}$ & 6, 7 \\
  \hline $C_{ij}$ [GPa] \cite{Shang} & $-2.0\,\%$ (0.01) & 23 $_{-2}^{+3}$ &  \\ 
  \hline \hline
 \end{tabular}
\end{table}

This statistical treatment makes several implicit assumptions. One of the most important premises is the use of a relative error over a constant offset. After all, for strictly positive quantities, such as $V_0$ or $T_m$, an invariable shift seems counterintuitive. The impact of such an error is much larger when the investigated property is small. In addition, when using relative systematic deviations, the difference from $\beta=1$ indeed matters. For the equilibrium volume ($p = 6 \cdot 10^{-11}$) and the bulk modulus ($p = 5 \cdot 10^{-4}$) the deviation from the bisector $x_{exp}=x_{th}$ is clearly significant. For other properties this is not always that obvious, but due to physical connections with $V_0$ and $B_0$, it is relevant to consider systematic deviations there as well.

\begin{figure}[t]
 \subfigure[]{\includegraphics[width=0.47\textwidth]{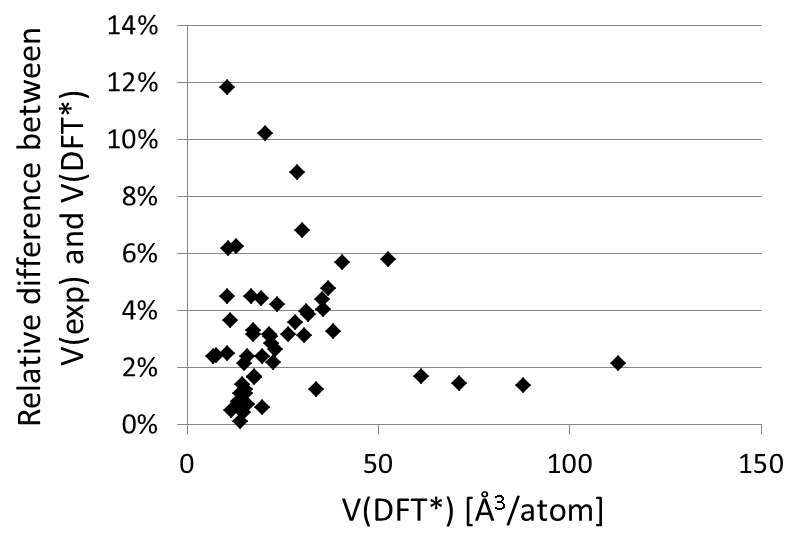} \label{figrelerror1}} \\ \subfigure[]{\includegraphics[width=0.47\textwidth]{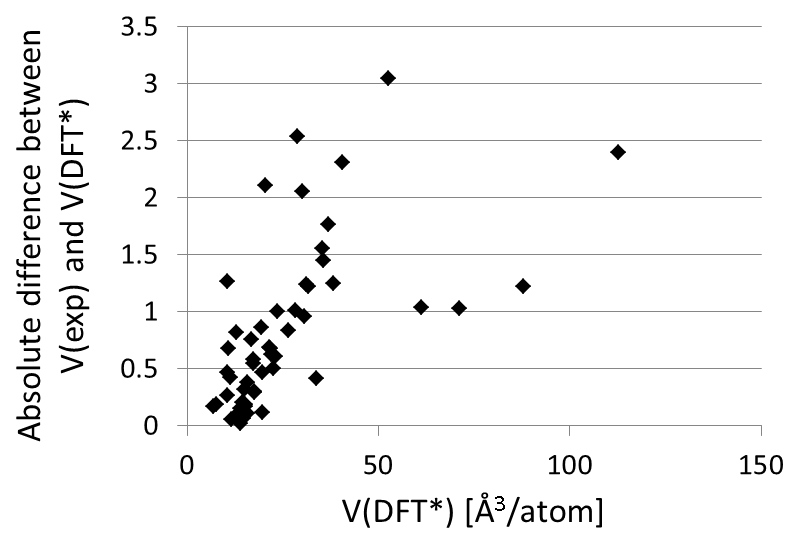} \label{figrelerror2}}
 \caption{Relative (a) and absolute (b) differences of the DFT regression data (DFT*) with respect to thermally corrected experimental values \cite{Villars,Bolz,Powell,Batchelder,Peterson,Losee,Sears,Swenson,Meyer} for the equilibrium volume \label{figrelerror}}
\end{figure}

\begin{table*}[t]
 \caption{Demonstration of the use of intrinsic errors (Tab.~\ref{tabaverror}) to improve DFT-PBE predictions and assess their reliability \label{taberrorproc}}
 \begin{tabular}{l|c|c|c}
  \hline \hline & $V_0$(W) [\AA$^3$/atom] & $B_0$(diamond) [GPa] &  $V_0$(GaAs) [\AA$^3$/atom] \\
  \hline PBE (bare) & 16.28 & 434.8 & 23.73 \cite{Haas} \\
  systematic deviation & 15.69 ($-3.6$\,\%) & 456.0 ($+4.9$\,\%) & 22.87 ($-3.6$\,\%) \\
  zero-point correction & 15.71 ($+0.02$) & 438.4 ($-17.6$) & 23.00 ($+0.13$) \cite{Haas} \\
  residual error bar & $\vt{15.7 \pm 1.1}$ & $\vt{438 \pm 15}$ & $\vt{23.0 \pm 1.1}$ \\
  \hline experiment (0\,K) & \vt{15.8} \cite{Haas} & \vt{443} \cite{Kittel} & \vt{22.5} \cite{Haas} \\
  \hline \hline
 \end{tabular}
\end{table*}

A second remark concerns the nature of the intrinsic residual error bars. The numbers in Tab.~\ref{tabaverror} were computed assuming a normal distribution for the random error $\epsilon$. This results in an absolute residual error bar. Indeed, a Pearson's $\chi^2$-test does not contradict the applicability of a normal distribution to the intrinsic random errors. A null hypothesis assuming a Gaussian distribution around zero yields a 1-sided $p$-value of 0.21 for the volume. Within the DFT community, however, relative error bars are often implicity assumed. As a consequence, small volumes are expected to be predicted much more accurately, for example. According to our analysis, the matter appears to be not that simple, as is shown in
Fig.~\ref{figrelerror}. In Fig.~\ref{figrelerror1}, the relative residual errors are plotted against the volume obtained by the least-squares fit. The overall decreasing trend suggests that the DFT errors are best described in terms of absolute error bars. 
Fig.~\ref{figrelerror2} displays the absolute residual errors instead. A rough linear correlation emerges, that implies that on the contrary a relative error bar is more appropriate. Both conclusions are compatible, by assuming a relative residual error bar of about 3\,\% for small to median volumes ($< 50$\,\AA$^3$/atom) and an absolute residual error bar of approximately $1.5$\,\AA$^3$/atom for larger volumes (with the exception of a few outliers\footnote{The fitting error is primarily correlated to the bond type. Although badly performing subsets of materials have already been excluded from the test set, some crystal types are described significantly better by DFT than others, especially those with strong bonds (covalent $p$-bonds, half-filled $d$-shell elements). Their strong bonds also lead to more compact structures, which explains the smaller errors for smaller volumes (Fig.~\ref{figrelerror2}). Other subsets of crystals with similar volumes perform worse, however, 
and give rise to the larger relative errors in Fig.~\ref{figrelerror1}.}). The number of data points is not sufficiently large for this finding to be really convincing, however. We therefore prefer to adopt the overall absolute error bar of $1.1$\,\AA$^3$/atom (Tab.~\ref{tabaverror}) as being valid for all volumes. Admittedly, for most small volumes the intrinsic residual error bars are then overestimated.

For any given material, the information in Tab.~\ref{tabaverror} can now be used to determine a meaningful estimate of a certain property. As an example we consider the bulk modulus of diamond. This material is not included in the test set (the ground-state crystal structure of carbon is graphite), but it can be assigned to subset 5, the high-coordination $p$-block compounds (similar to Si, Ge, and Sn). Subset 5 does not belong to one of the excluded subsets for $B_0$, and the intrinsic error for the bulk modulus prediction of diamond should therefore be representative. A bare VASP-PBE computation yields $434.8$\,GPa. When taking into account that PBE bulk moduli are systematically too small by $4.9$\,\% (Tab.~\ref{tabaverror}), this value increases to 456.0\,GPa. By means of Eqs.~(\ref{dB2}), (\ref{dV2}), and (\ref{eqThetaD}) zero-point corrections ($-17.6$\,GPa) are added back in, yielding $438.4$\,GPa. Using the appropriate intrinsic residual error bar of 15\,GPa (Tab.~\ref{tabaverror}), the final result 
becomes $438 \pm 15$\,GPa. This is the most accurate DFT-PBE prediction of the experimental bulk modulus at 0\,K, including an error bar on the computed value. In comparison, the experimental value extrapolated to absolute zero (Eqs.~(\ref{dB1}) and (\ref{eqdV1b})) amounts to 443\,GPa \cite{Kittel}. This number remains neatly within the error bar and is indeed closer to the regression-corrected bulk modulus than to the bare DFT value. A similar procedure can be used for all properties in Tab.~\ref{tabaverror}. Tab.~\ref{taberrorproc} offers a few more examples.

Tab.~\ref{tabaverror} is based on elemental solids only. One has to verify that the results from this statistical analysis are transferable to multicomponent materials. A good test case in that respect is the collection of thirty-one binary compounds for which Haas \emph{et al.} \cite{Haas, Haaserratum} calculated lattice parameters by means of PBE. When we take their DFT results, the experimental volume falls outside the confidence interval for seven crystals. In all of these cases, the PBE volume is too large. By taking into account the systematic overestimation of 3.6\,\%, however, only in two cases the experimental value exceeds the predicted range. Both of these compounds are ionically bound (NaF and NaCl), a bond type that has not been considered in any of the proposed subsets (Fig.~\ref{figgroups}). Although the multicomponent test set may therefore not contain all materials types and although it only relates to the atomic volumes, these results already strongly support the transferability of our 
error estimates.

For one example from Haas \emph{et al.}, GaAs, the intrinsic error contributions are listed in Tab.~\ref{taberrorproc}. It illustrates that the systematic deviation really matters if it is the goal to get as close as possible to the experimental value. This also appears from the mean absolute difference between experiment and the 31 theoretical predictions by Haas \emph{et al.}. When one does not apply the relative deviation from Tab.~\ref{tabaverror}, this number amounts to $0.72 \, \sigma_V$ ($0.78$\,\AA$^3$/atom), while for the regression values it is only $0.37 \, \sigma_V$ ($0.40$\,\AA$^3$/atom).

\subsection{Agreement with experiment \label{secerror}}

\subsubsection{Errors per materials type}

Because Tabs.~\ref{taberrors1}-\ref{taberrors2} are shaped like the periodic table, the color code immediately allows to single out the areas where PBE breaks down. It leads to a number of subsets (Fig.~\ref{figgroups}) which can be eliminated from the test set. They are listed in Tab.~\ref{tabaverror}. For these elements, PBE performs significantly worse than usual, mostly because some key physical phenomenon is not described (well) by the functional. In this subsection these localized error zones are discussed in more detail, as well as the mechanisms on which they are based.

A first, well-known example of the failure of PBE is the class of \textbf{dispersion-governed compounds}. Although some more advanced DFT approaches address this issue specifically \cite{Grimme, vdwdft}, regular GGA functionals do not describe London forces. This translates into a decreased cohesion, and hence an inflated volume and underestimated bulk modulus (see also Sec.~\ref{secgeneralerrors}).

Although London forces have been demonstrated to play a role in other structures as well \cite{vdwas,vdwnoble,vdwp}, the most important crystals that suffer from this shortcoming, belong to the nonmetals (subsets 7 and 8). They include the noble gases, the dimeric crystals, graphite, and sulfur. In these materials the London dispersion interaction governs the bonding between atoms, diatomic molecules, graphene sheets, and 8-membered rings, respectively. Nevertheless it is essential to realize that both the element type and the crystal structure contribute to the importance of dispersion. It is perfectly plausible that a certain element behaves badly in structure A, while there are no problems when it assumes structure B. This can be illustrated nicely by means of carbon. The dispersion forces between the graphene layers in graphite give rise to a large discrepancy between DFT and experiment. Diamond on the other hand follows the same behavior as neighboring (semi)metallic 
elements or even outperforms them (Tab.~\ref{tabC}).

\begin{table}[b]
 \caption{Residual errors (of the VASP-PBE regression results with respect to the zero-kelvin extrapolated experimental values \cite{Kittel,Greenwood,Tohei,Villars,graphite}) for two allotropes of carbon \label{tabC}} \vspace{0.3cm}
 \begin{tabular}{l|c|c|c}
  \hline \hline & $\Delta E_{coh}$ [kJ/mol] & $V_0$ [\AA$^3$/at] & $B_0$ [GPa] \\
  \hline graphite & 39 (5\,\%) & 3.2 (39\,\%) & 54 (97\,\%) \\
  diamond & 13 (2\,\%) & 0.1 (2\,\%) & 5 (1\,\%) \\
  \hline \hline
 \end{tabular}
\end{table}

For the molecular crystals (subset 7) the PBE cohesive energy is larger than the experimental value \cite{suppl}, contrary to the expectation. This is due to the overestimation of the intramolecular bond strength (see e.g. Lany \cite{Lany} and Tab.~II of Paier \emph{et al.} \cite{g2-1}), which covers up any influence of the lack of dispersion. Elastic properties on the other hand are in most cases not affected by intramolecular effects and show a similar behavior as for the remaining nonmetals.

The \textbf{magnetic materials} (subset 3) stand out as well, predominantly with respect to $\Delta E_{coh}$. Although the use of the generalized gradient approximation and a correct atomic reference state have already reduced the gap between theory and experiment substantially \cite{Philipsen1996}, the remaining difference cannot be neglected. Current GGA functionals are not able to describe magnetic compounds very well. Manganese illustrates this nicely. Its intricate magnetic state \cite{Hobbs} has been approximated by assuming only collinear magnetism, but this does not explain the observed differences. The cohesive energy, for example, would be higher in its correct ground state, leading to an even more pronounced deviation from experiment. An explanation is found with Singh and Ashkenazi \cite{Singh}, who noticed that GGAs overestimate the magnetic energy. This is caused by the increased number of degrees of freedom in spin-polarized systems (two spins), while the number of physical relations the GGA 
must fulfill stays the same. 

The discrepancies between theory and experiment are not caused by the DFT functional alone, however. For some magnetic elements the applied thermal extrapolations are no longer valid, because of phase transition effects in the vicinity of the Curie or N\'eel temperature. Experimental chromium is a good example, displaying large magnetic distortions of the thermal expansion coefficient near 311\,K \cite{Roberts}. A relation as simple as that of Eq.~(\ref{eqdV1b}) cannot capture these complex underlying phenomena.

The transition metals with (nearly) full $d$ shells sometimes deviate from experiment as well (subset 4). The effects are smaller than for the previous two classes of materials, but they are unmistakably present, especially in terms of the cohesive energy and the elastic constants. One can attribute this phenomenon to \textbf{electronic correlation}. For Zn, Cd, and Hg a full-fledged many-body treatment has indeed convincingly shown the influence of $d$ electron correlation on $\Delta E_{coh}$ and the potential energy landscape \cite{Gaston, Wedig, Gaston2010}. Data in Tabs.~\ref{taberrors1}-\ref{taberrors2} even imply that similar (but smaller) effects show up in other elements at the end of the $d$ block, such as in Pd, Ag, Pt, and Au. In noble metals, dispersion phenomena play an important role too \cite{vdwnoble}, however, and it is not immediately clear how much of the remaining discrepancy can be attributed to electronic correlation. Since the influence of correlation in these 
elements appears limited, only Cd an Hg have been assigned to subset 4.

It seems that at the end of the $d$ block, the high number of localized $d$ electron pairs in combination with a small interatomic distance and a close-packed environment enhances correlation effects. Moreover, Philipsen and Baerends \cite{Philipsen1996} suggest that at the very beginning and the very end of the 3$d$ transition metals the GGA exchange energy drops, causing the electron correlation to gain importance. Any fortuitous cancellation of errors between exchange and correlation has therefore disappeared for the transition metals at the border of the $d$ block.

These correlation effects appear to be of a mainly anisotropic nature, since all elastic constants $C_{ij}$ are affected, but the deviations of $V_0$ and $B_0$ are less pronounced. This is also suggested by Wedig \emph{et al.} \cite{Wedig} for Zn and Cd, where a different interlayer and intralayer behavior is observed.

\textbf{Relativistic effects} are expected to strongly influence heavy elements. VASP therefore makes use of the scalar-relativistic Kohn-Sham equations by default \cite{Koelling}. The major remaining contribution is due to spin-orbit coupling. However, it is shown by Philipsen and Baerends \cite{Philipsen} that this does not change physical properties substantially. Only for gold and bismuth a distinct change is reported, but without closing the gap between theory and experiment entirely. The remaining difference for Au is primarily due to correlation and dispersion effects, as was already suggested above. For the 6$p$ elements on the other hand spin-orbit coupling really plays an important role. Some key properties for the $5d$ and $6p$ compounds have been calculated, both with and without spin-orbit coupling (Tab.~\ref{tabrel}). It is immediately clear that, starting from the end of the 5$d$ block, a spin-orbit treatment becomes indispensable. Hence, for all 5$d$ and 6$p$ elements this contribution has 
been 
included, 
except for $C_{ij}$ \cite{Shang}.

\begin{table}[t]
 \caption{Relative residual errors (of the VASP-PBE regression results with respect to the zero-kelvin extrapolated experimental values \cite{Kittel,Villars,Swenson}) for Ag and the $5d$ and $6p$ materials, both with (SO) and without spin-orbit coupling (n-SO) \label{tabrel}} \vspace{0.3cm}
 \begin{tabular}{l|cc|cc|cc}
  \hline \hline & \multicolumn{2}{c|}{$\Delta E_{coh}$} & \multicolumn{2}{c|}{$V_0$} & \multicolumn{2}{c}{$B_0$} \\
  \hline & n-SO & SO & n-SO & SO & n-SO & SO \\
  \hline Ag & 16\,\% & 16\,\% & 3\,\% & 3\,\% & 11\,\% & 11\,\% \\
  \hline Lu & 7\,\% & 8\,\% & 3\,\% & 3\,\% & 18\,\% & 18\,\% \\
  Hf & 1\,\% & 3\,\% & 3\,\% & 3\,\% & 2\,\% & 2\,\% \\
  Ta & 1\,\% & 2\,\% & 2\,\% & 2\,\% & 1\,\% & 0\,\% \\
  W & 0\,\% & 1\,\% & 1\,\% & 1\,\% & 3\,\% & 5\,\% \\
  Re & 3\,\% & 2\,\% & 2\,\% & 1\,\% & 3\,\% & 1\,\% \\
  Os & 1\,\% & 3\,\% & 0\,\% & 1\,\% & 1\,\% & 4\,\% \\
  Ir & 4\,\% & 0\,\% & 0\,\% & 0\,\% & 0\,\% & 2\,\% \\
  Pt & 7\,\% & 9\,\% & 1\,\% & 1\,\% & 9\,\% & 11\,\% \\
  Au & 22\,\% & 19\,\% & 4\,\% & 3\,\% & 21\,\% & 19\,\% \\
  Hg & 75\,\% & 69\,\% & 29\,\% & 22\,\% & & \\
  \hline Tl & 6\,\% & 20\,\% & 9\,\% & 7\,\% & 27\,\% & 25\,\% \\
  Pb & 44\,\% & 4\,\% & 4\,\% & 4\,\% & 10\,\% & 18\,\% \\
  Bi & 15\,\% & 5\,\% & 2\,\% & 5\,\% & 17\,\% & 26\,\% \\
  Po & 59\,\% & 8\,\% & 2\,\% & 3\,\% & 73\,\% & 30\,\% \\
  Rn & 83\,\% & 81\,\% & & & & \\
  \hline \hline
 \end{tabular}
\end{table}

\subsubsection{Errors per property \label{secgeneralerrors}}

The previous section shows that PBE is not `complete': some features just cannot be described by a simple GGA functional. One can exclude the affected materials (outliers) beforehand, however, and limit the analysis to those cases where PBE should perform well. The intrinsic errors from Tab.~\ref{tabaverror} are applicable to these crystals. Tab.~\ref{tabaverror} then shows that the PBE error estimates largely depend on what property is considered. The behavior of the residual error bar and the systematic deviation from experiment can be traced back to both the functional and the numerical determination of that particular property. Nevertheless, it is important to note that, although the overall error estimate can be linked to theoretical aspects, the correspondence to experiment for a single compound depends on the experimental accuracy as well. This is especially true for higher-order properties (such as the elastic constants 
and their derivatives), which are generally measured at a lower precision than those from (quasi)direct measurements (the lattice constants or the cohesive energy, for example). This is illustrated by the sometimes large spread on the data in Knittle's overview of $B_1$ values \cite{Knittle}.

From a computational viewpoint, however, the equilibrium \textbf{volumes} offer the best results among all considered quantities. After eliminating the outliers (listed in Tab.~\ref{tabaverror} and represented in Fig.~\ref{figregV} by open symbols), an almost perfect correlation is obtained. Even so, the regression line does not coincide with the first quadrant bisector. Tab.~\ref{tabaverror} shows that the cell volumes are consistently too large by approximately 4\,\%. This deviation is a well-known property of any GGA \cite{Ozolins}, including PBE. It originates in a systematic underestimation of the bond strength (underbinding), resulting in slightly larger volumes. More particularly, GGAs favor inhomogeneous systems, with large (reduced) density gradients. Small unit cells, which have a more evenly distributed electron density, are therefore energetically less preferable. This phenomenon especially affects open structures, where the high-gradient tails of the valence electron orbitals become 
non-negligible \cite{Haas2}. 

Immediately linked to this observation is the underestimation of the \textbf{bulk modulus}. More weakly bound structures will be more easily compressible, leading to smaller $B_0$ values. Just like the too large predicted volumes, it is common behavior for GGA functionals \cite{Ozolins}.  On the other hand, PBE bulk moduli are predicted with a larger uncertainty than the volumes. The intrinsic residual error remains in most cases below 10 to 15\,\%, however (Tab.~\ref{taberrors1}). The magnitude of this difference is mostly due to the sensitivity of the $E(V)$ curvature. 

Since $B_0$ and the other \textbf{elastic constants} are closely related, the intrinsic errors with respect to the $C_{ij}$ parameters are of a comparable scale. The bulk moduli are larger on average, which leads to slightly smaller relative errors (Tabs.~\ref{taberrors1} and \ref{taberrors2}). However, a good correlation is found in both cases, with a similar value of $r$ for the elastic constants (Fig.~\ref{figreg2}) and the bulk moduli (Fig.~\ref{figregB}).

For the \textbf{cohesive energy} PBE yields very good results as well. The intrinsic error bar of 30\,kJ/mol is of the same order of magnitude as the rms error found by Lany \cite{Lany} for PBE heats of formation of semiconductors and insulators (0.24\,eV/atom). It is therefore representative for PBE energy differences between chemically different compounds. As mentioned before, for similar systems the intrinsic residual error bar is much smaller \cite{Hautiererror}. This also explains the success of evolutionary algorithms. They are based on energy differences of the order of a few meV per atom \cite{USPEX, genetic1, genetic2, genetic3}, but some results have already been confirmed experimentally nevertheless \cite{Ma}.

Contrary to $V_0$ and $B_0$ there is now no systematic under- or overestimation compared to experiment. The typical underbinding of GGA does not show as conclusively in $\Delta E_{coh}$. This is due to the magnetic materials and the molecular crystals. As mentioned before, GGA functionals bias solutions towards magnetism for the former and overestimate the intramolecular contribution in the latter. In both cases this causes the cohesive energy to oppose the dominating trend. Without the crystals from subsets 3 and 7 in the test set, the cohesive energies would have been underestimated by $2\,\%$ instead (p-value of $0.008$). This behavior is in accordance with the expected underbinding of GGA.

Since the \textbf{bulk modulus derivative} $B_1$ is a higher-order parameter than $B_0$, the errors are expected to be one order worse as well. Although this is certainly the case, eliminating the outliers substantially improves the results (Fig.~\ref{figregBP}). However, even when they are removed, the resulting correlation coefficient ($r=0.849$) remains significantly lower than for any other property already discussed. 

$B_1$ appears to be overestimated with respect to experiment. This systematic deviation is significant, although the p-value may not show it conclusively (Tab.~\ref{tabaverror}). It is again caused by GGA underbinding. As mentioned before, large volumes are favored due to their substantial density gradients. GGA hence lowers the energies of bigger cells most and straightens out the equation of state. This causes the $E(V)$ line to alter its decreasing curvature even more rapidly, increasing the rate of change of the bulk modulus with pressure (and volume), $B_1$. It also explains the deviating behavior of crystals with a low coordination, such as the molecular crystals. In these compounds the tails of the electron wave functions dominate the interstitial space, leading to considerable density gradients. The increase of the sensitive parameter $B_1$ is then enhanced even further.

\section{Numerical errors \label{seccode}}

The previous section describes the intrinsic PBE errors for five different properties, based on a statistical treatment. They are used in a protocol which allows experimentalists and theoreticians to correct the bare DFT-PBE values for the observed systematic deviation from experiment and which quantifies the uncertainty on the obtained predictions (Tab.~\ref{taberrorproc}). A prerequisite for such a protocol is that different DFT implementations provide the same predictions: using different algorithms to solve the same (Kohn-Sham) equations should ideally lead to identical solutions. In practice, different amounts of noise are inevitably introduced in the predictions, even when numerical convergence has been achieved for each individual code. This scatter is due to several aspects of the solution algorithm. It can be due its nature (e.g. the kind of basis set or the frozen-core approximation), its specific ingredients (e.g. the chosen pseudopotential) 
or its use of particular routines for 
standard tasks (e.g. Fourier transform routines). In order to guarantee the 
reproducibility of the intrinsic errors in Tab.~\ref{tabaverror}, one needs to examine to what extent these issues affect the DFT computations. Once again, a reliable 
benchmark can be established using the ground-state elemental crystals.

The current section describes a procedure to express the difference between predictions from independent solid-state DFT approaches in a quantitative way, yielding a numerical error estimate. It will be used to examine differences between the PAW and APW+lo method, representing both methods by suitable mainstream codes (see further). However, the difference between codes can be attributed to other aspects as well, as was mentioned earlier. The influence of standard task routines is most likely small, but we will also use two PAW codes with different PAW atomic potentials, which can have quite drastic effects on the DFT results. All codes are therefore considered here with their recommended potentials. These can be thought of as representative for the quality of the investigated code. Although it is not our primary goal here, the procedure that will be described can also be used to select better performing PAW atomic potentials. The same holds for pseudopotentials in the case of plane-wave 
codes.

\subsection{Test set preparation}

The present implementation assessment starts from one reference code, the all-electron program WIEN2k \cite{wien2k} (version 11.1). It uses the APW+lo basis set \cite{Sjoestedt, Madsen}, which is considered to be a standard for the numerical accuracy of solid-state DFT. WIEN2k predictions can therefore be considered to yield the exact results for a given functional, as long as numerical accuracy is achieved \cite{Haas} (large basis set and dense $k$-mesh, see Supplementary Material for more details \cite{suppl}). Two codes are compared to this reference code: VASP \cite{vasp, vasp2} (version 5.2.2) and GPAW \cite{ASE, gpaw, gpaw2} (version 0.8.0), both using the PAW method \cite{Bloechl}. GPAW calculates all wave functions, densities and potentials as grid-based quantities, while VASP uses a plane-wave basis set. All calculations with these programs are performed by means of the potentials recommended by the respective developers: the 2010 recommended PAW potentials \cite{suppl} 
for 
VASP and the 0.6 atomic set-ups for GPAW. Detailed computational parameters are summarized in the Supplementary 
Material \cite{suppl}.

For reasons of uniformity and comparability the same PBE functional has been selected for all three codes. It is used in a protocol that seeks to evaluate a particular DFT approach in an easily reproducible manner. The VASP-optimized ground-state crystal (Sec.~\ref{secpropcomp}) serves as a starting point for each computation and from a 7-point equation of state ($0.94\,V_0$ to $1.06\,V_0$) the properties of interest ($E_0$, $V_0$, $B_0$, and $B_1$) are extracted. All geometries are kept frozen (the cell shape and relative atomic positions are kept fixed at their initial values), instead of allowing for relaxation changes. This not only lowers the computational load, it also restricts the code evaluation to the implementation of DFT-PBE itself. Indeed, the task of optimizing the cell shape or internal positions belongs to another computational layer, on top of the task of solving the DFT equations for a given rigid geometry. This section aims to examine how different implementations compare 
with respect to the DFT-PBE 
procedure only. It does not intend to study how close every individual approach comes to experiment.

In the same spirit some other modifications of the hitherto employed test set have been made. All calculations have been limited to the scalar-relativistic part (using the Koelling-Harmon approach \cite{Koelling}). By neglecting the spin-orbit contribution, an additional secondary algorithm implementation is avoided. The computational procedure also becomes more uniform this way, since all elements are now treated on equal terms. Because no spin-orbit coupling is added to the system's Hamiltonian, it suffices to use non-spin-orbit geometries as a starting point for the 7-point equation of state.

A simplified unit cell has been selected for Mn and S as an additional means of lowering the computational effort. Manganese is treated in an antiferromagnetic fcc phase (space group 225, cF4), while for sulfur the $\beta$ Po phase is imposed (space group 166 or hR3). These geometries are physically relevant, as they can be found in the Mn and S phase diagram respectively \cite{Rapoport, Luo}.  

All other elements have been kept at the structure previously optimized by VASP (Sec.~\ref{secpropcomp}), in order to conserve the large diversity of the input set. The CIF files for all crystals in this code benchmark set are available in the Supplementary Material \cite{suppl}.

\subsection{Agreement between implementations}

\begin{figure}[b]
 \includegraphics[width=\columnwidth]{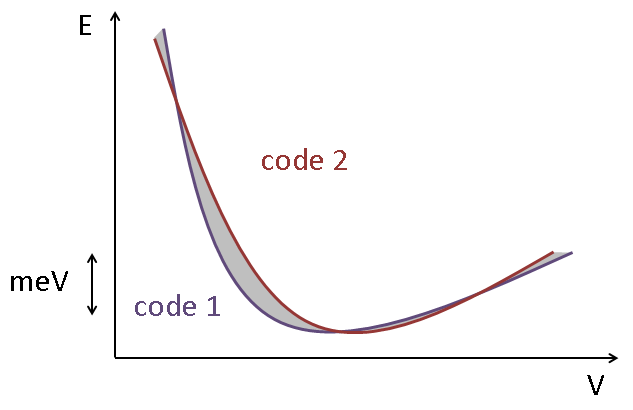}
 \caption{(Color online) The EOS parameters can differ significantly, while the $E(V)$ curves themselves are very similar. In that case the area between the two functions is a better indicator of the overall deviation \label{figdelta}}
\end{figure}

The procedure mentioned above results in a large collection of numbers for each code (71 elements $\times$ 4 properties). It is not convenient to compare them directly, however. Because of the different units involved, a coherent approach would require the use of relative deviations. Tabs.~\ref{taberrors1}-\ref{taberrors2} on the other hand show that each property corresponds to a different magnitude of relative error. This scale is mainly determined by the computational procedure and therefore does not alter substantially when shifting from a code-experiment comparison to an intercode assessment. A single numerical error value, expressing the difference between two particular DFT methods by means of one number, can be obtained by applying a weighed average. As all properties of interest depend on the equation of state, it is most straightforward to compare the $E(V)$ curves produced by different approaches directly. The dispersion-governed compounds illustrate this 
strategy well. Since 
their $E(V)$ 
curves 
are very shallow, small deviations in the bulk modulus will inflate the relative error considerably. However, the equations of state as such can be very similar, the two curves at no point differing by more than a few meV per atom (Fig.~\ref{figdelta}). For that reason a numerical error estimate $\Delta$ is defined as follows:
\begin{equation}
 \Delta = \left\langle \sqrt{\frac{\int \Delta E^2 (V) \, \mathrm{d}V}{\Delta V}} \right\rangle \label{eqdelta}
\end{equation}
In other words, the rms energy difference between the $E(V)$ curves of these particular programs is averaged over all elemental crystals. $\Delta$ hence provides an intuitive measure of the energy distance between equations of state.

Because different codes sometimes employ different reference energies $E_0$, depending on the concept, all equations of state are set to zero at their equilibrium volume. An alternative solution would entail the calculation of cohesive energies, in order to provide a common reference for the equilibrium energy. However, not all programs allow for an easy manipulation of the electronic configuration of atoms. Moreover, the computational load would increase considerably.

The computation of $\Delta$ can be automated quite easily. The fitted Birch-Murnaghan equation allows Eq.~(\ref{eqdelta}) to be written in an analytical form. Only $V_0$, $B_0$, and $B_1$ are then needed for both codes under investigation. The resulting expressions have been added in the appendix for convenience. The WIEN2k data necessary for a code comparison have been provided in the Supplementary Material \cite{suppl}.

The interval of integration is linked to the reference data. In view of how the $E(V)$ parameters are determined, the intercode difference is to be integrated between $V_{0,WIEN2k} \pm 6\,\%$. $\Delta V$ hence corresponds to $0.12 \, V_{0,WIEN2k}$. By definition $\Delta\text{(APW+lo)}_\text{(WIEN2k)}$ becomes zero.

The rms energy differences between the equations of state predicted by APW+lo$_\text{(WIEN2k)}$ and PAW$_\text{(VASP)}$, or APW+lo$_\text{(WIEN2k)}$ and PAW$_\text{(GPAW)}$, are represented in Tab.~\ref{tabdelta}. They show that most critical elements are characterized by approximately half-filled $d$ levels. Such numerical errors can amount to up to 8.3\,meV/atom for PAW$_\text{(VASP)}$ (Tc) and 20.9\,meV/atom for PAW$_\text{(GPAW)}$ (Ru). This agrees with physical intuition, because these crystals are among the least compressible. Their equations of state are very steep, and relatively small modifications of the parameters can strongly change the energy. The least sensitive elements are for the same reason located near the alkali metals and the noble gases (0\,-\,0.7\,meV/atom numerical error) (see Tab.~\ref{tabmeaningdelta}). Only in comparison to experiment the latter group of materials stands out, but this is because PBE grossly overestimates the rare gas 
volumes.

\begin{table}[t]
 \caption{Comparison between codes for two extreme situations: large (Tc an Ru) and small (Ar) numerical errors $\Delta_i$. $V_0$ is given in \AA$^3$/atom, $B_0$ in GPa, and $\Delta_i$ in meV/atom. $B_1$ is dimensionless \label{tabmeaningdelta}} \vspace{0.3cm}
 \begin{tabular}{l|c|c|c|c|c}
  \hline \hline & & $V_0$ & $B_0$ & $B_1$ & $\Delta_i$ \\
  \hline Tc & APW+lo$_\text{(WIEN2k)}$ & 14.47 & 301.4 & 4.56 & 0 \\
   & PAW$_\text{(VASP)}$ & 14.60 & 298.5 & 4.55 & 8.3 \\
  \hline Ru & APW+lo$_\text{(WIEN2k)}$ & 13.81 & 315.4 & 4.96 & 0 \\
   & PAW$_\text{(GPAW)}$ & 14.09 & 310.9 & 4.87 & 20.9 \\
  \hline Ar & APW+lo$_\text{(WIEN2k)}$ & 52.21 & 0.7 & 7.84 & 0 \\
   & PAW$_\text{(VASP)}$ & 52.65 & 0.8 & 7.35 & 0.1 \\
   & PAW$_\text{(GPAW)}$ & 52.66 & 0.8 & 3.27 & 0.1 \\
  \hline \hline
 \end{tabular}
\end{table}

\begin{table*}[p]
 \caption{(Color online) Rms energy differences $\Delta_i$ between the equations of state predicted by APW+lo$_\text{(WIEN2k)}$ and PAW$_\text{(VASP)}$ (green), APW+lo$_\text{(WIEN2k)}$ and PAW$_\text{(GPAW)}$ (red), and experiment and PAW$_\text{(VASP)}$ (blue) for the ground-state elemental crystals. All values are expressed in meV per atom. The darkest shades correspond to the largest errors. The average numerical error $\Delta$ is shown for each code at the header of the table \label{tabdelta}}
 \vspace{1ex}
 \includegraphics[width=\textwidth]{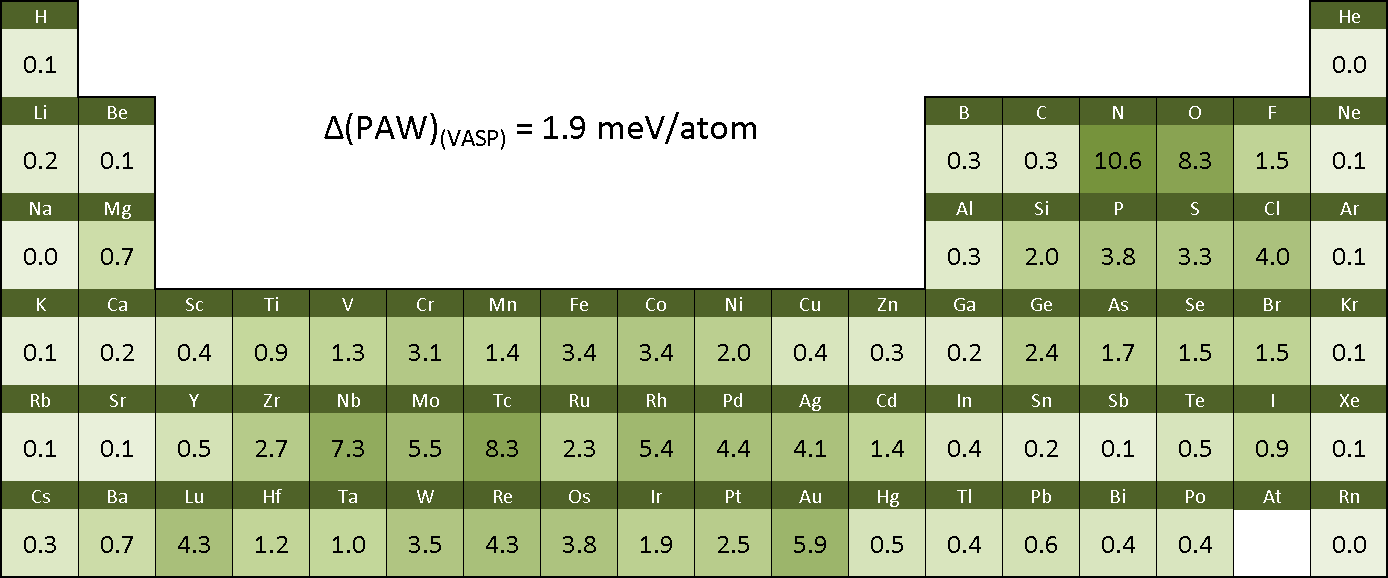} \\ \vspace{0.5cm}
 \includegraphics[width=\textwidth]{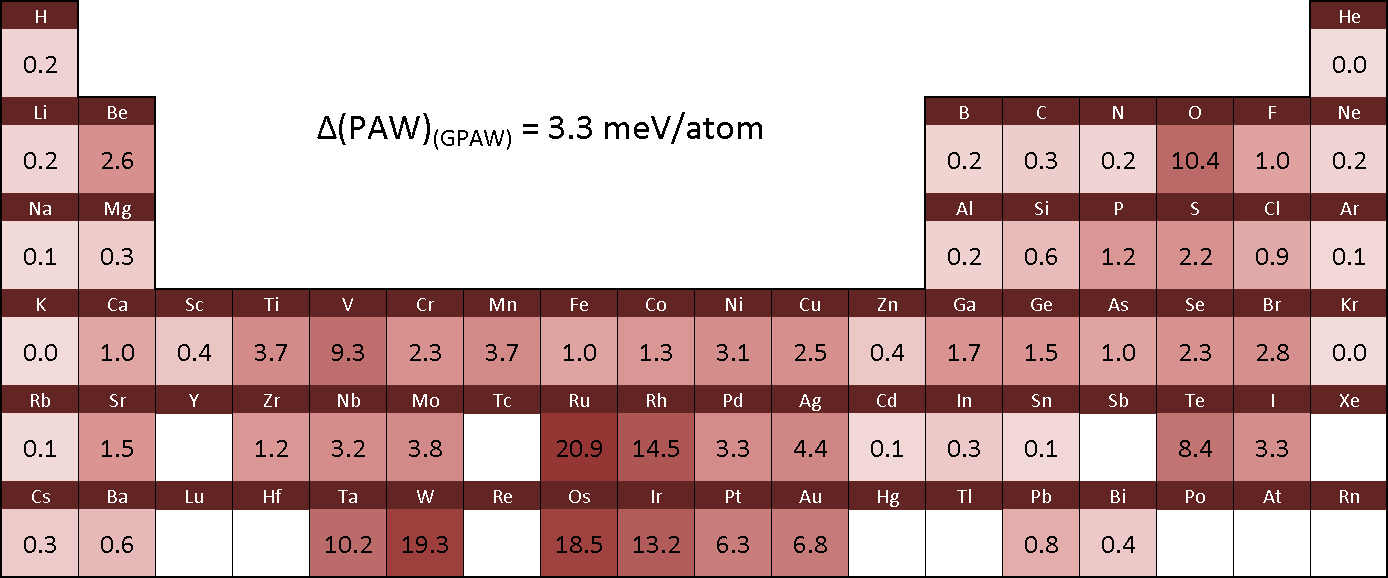} \\ \vspace{0.5cm}
 \includegraphics[width=\textwidth]{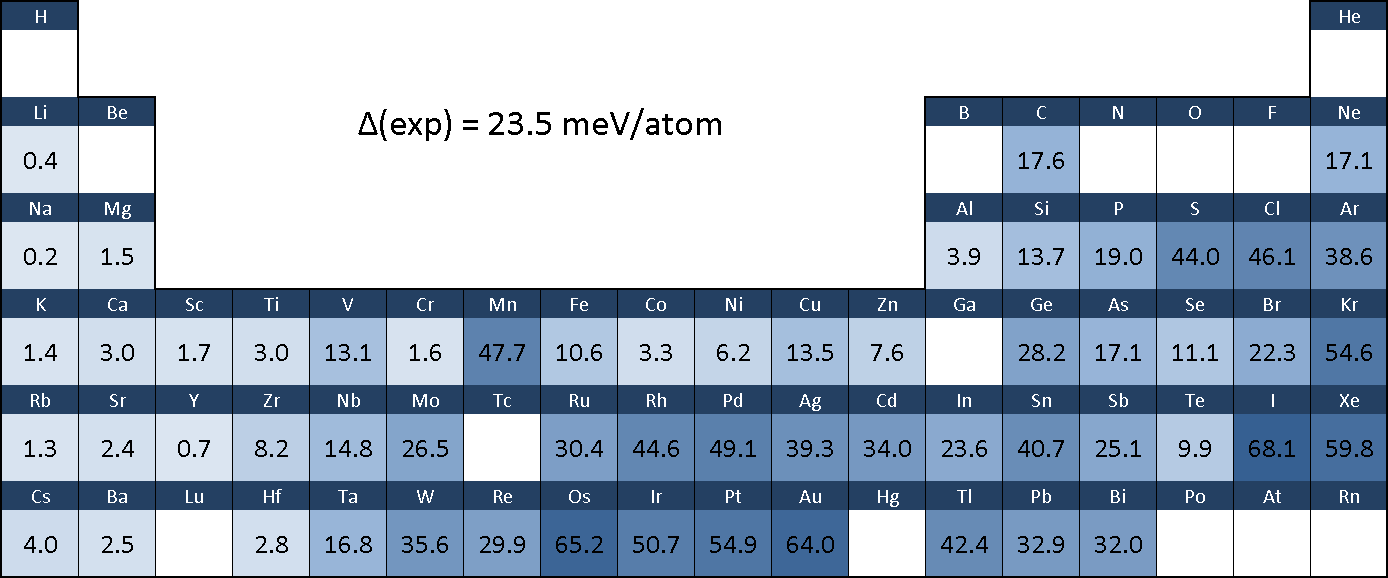}
\end{table*}

\begin{figure}[t]
 \includegraphics[width=\columnwidth]{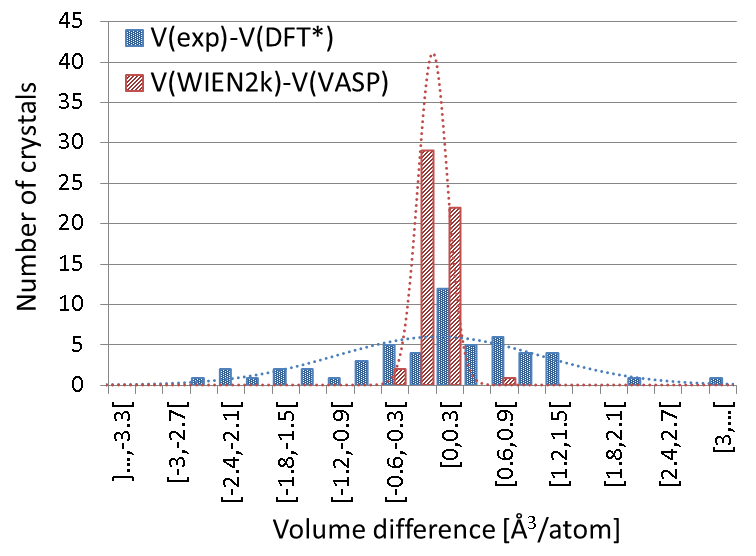}
 \caption{(Color online) Intrinsic (PBE regression versus experiment) and numerical errors (PAW$_\text{(VASP)}$ versus APW+lo$_\text{(WIEN2k)}$) for the equilibrium volume of the ground-state elemental crystals, using the subsets of elements that have been shown to perform well for PBE (see Tab.~\ref{tabaverror}). A normal distribution has been fitted to both data sets (dotted line) \label{fighistcode}}
\end{figure}

When averaging the numbers in Tab.~\ref{tabdelta} over all elements, the numerical error of each DFT approach can be determined for the given set of recommended PAW potentials. $\Delta$ is 1.9\,meV/atom for PAW$_\text{(VASP)}$, while for PAW$_\text{(GPAW)}$ it is 3.3\,meV/atom. This agreement between implementations is an order of magnitude better than the difference with experimental results. To show this, a similar energy difference between DFT-PBE and experiment is computed. It uses experimental values as the reference situation, while the method under test is the full-fledged version of PAW$_\text{(VASP)}$. This means that the $E(V)$ parameters have been taken from Tabs.~\ref{taberrors1} and \ref{taberrors2}. The deviations per element are presented in Tab.~\ref{tabdelta}, leading up to a $\Delta$-factor of 23.5\,meV/atom. This difference in magnitude can also be observed with the $E(V)$ characteristics themselves. Fig.~\ref{fighistcode} 
shows the distribution of volume errors between two codes and with respect to experiment. Again, 
the spread is 
much 
larger in the latter case.

$\Delta$(PAW)$_\text{(VASP)}$ does not change noticeably when the number of elements is reduced to that of PAW$_\text{(GPAW)}$. This shows that GPAW and VASP, while both using the same PAW method, do not produce entirely identical results. This variation most likely originates in the different quality of the atomic potentials and the different type of basis functions used. However, in comparison to experiment, the differences are negligible. The intrinsic residual error bars and regression slopes provided in Tab.~\ref{tabaverror} can therefore be applied to DFT-PBE results, irrespective of which approach was used to calculate them.

This comparison of three DFT implementations can easily be extended. Ideally every solid-state DFT approach should be tested in the same way, and have its $\Delta$-value computed. As such tests are preferably performed by specialists in the individual codes, all input CIF files have been made available in the Supplementary Material \cite{suppl}, as well as some ready-made post-processing scripts. In addition, the ASE developers have implemented a framework for performing the necessary calculations \cite{Dulak, ASE}. On the CMM website \cite{molmod} an updated overview will be maintained of all $\Delta$-factors reported to us. Such information not only provides insight into the reproducibility of the intrinsic errors of Tab.~\ref{tabaverror}, but can also guide users to select a method for a specific task, at least as far as accuracy of energy-versus-volume relations is concerned.

\section{Conclusions \label{secconclusion}}

Using the ground-state elemental crystals as a test set, DFT-PBE computational errors have been reviewed. Errors intrinsic to the functional were quantified for five materials properties, describing energetic ($\Delta E_{coh}$) and elastic ($V_0$, $B_0$, $B_1$, $C_{ij}$) quantities. They explain the deviation of DFT predictions from experiment. Numerical errors, due to the implementation of the DFT scheme into a computer code, were studied for the PAW method (VASP and GPAW), and were expressed with respect to the reference APW+lo method (WIEN2k). Both types of errors have been discussed for PBE, one of the most widely applied functionals in solid-state DFT. The results are expected to be representative of GGA in general.

Each of the five properties has been assessed with respect to the ground-state elemental crystals. The correspondence to experiment has been analyzed statistically, leading to a decomposition of the intrinsic error into systematic deviations and residual error bars. These intrinsic errors have been shown to agree with some generally known GGA traits. The typical underbinding of GGA has been reproduced and quantified, for example. Tab.~\ref{tabconcl} presents a summary of our results, as well as a similar analysis for other properties that are available from data sets in the literature. Contrary to Tab.~\ref{tabaverror}, however, Tab.~\ref{tabconcl} presents systematic deviations in terms of the experimental value ($(x_{th}-x_{exp})/x_{exp} = 1/\beta-1$). This expresses more intuitively to what extent DFT varies from experiment: a $1/\beta-1$ of $+1$\,\% for example means that PBE overestimates the experimental result by $1$\,
\%.

\begin{table*}[t]
 \caption{Systematic deviations $1/\beta-1$ and residual error bars for DFT-GGA predictions compared with experiment (intrinsic errors) and between codes (numerical errors). $\Delta$ represents the average rms energy difference between the equations of states of two codes. All data were (re)analyzed in the present study, except for the error bar for $\Delta E_{evol}$, which is based on a proof of principle (see Sec.~\ref{secgeneralerrors}). Subsets of materials to which the error estimates do not apply are mentioned explicitly \label{tabconcl}} \vspace{0.3cm}
\begin{tabular}{l|c|c|c}
  \hline \hline \multicolumn{4}{c}{Intrinsic errors (this work, VASP-PBE)} \\
  \hline  & systematic deviation & residual error bar & not applicable to: \\
  \hline $\Delta E_{coh}$ [kJ/mol] & \phantom{1}$-0.0\,\%$ & \phantom{2}30\phantom{.00} & mostly correlation; pure dispersion \\
  $V_0$ [\AA$^3$/atom] & \phantom{1}$+3.8\,\%$ & \phantom{28}1.1\phantom{0} & mostly correlation; dispersion \\
  $B_0$ [GPa] & \phantom{1}$-4.7\,\%$ & \phantom{2}15\phantom{.00} & dispersion \\
  $B_1$ [--] & \phantom{1}$+5.0\,\%$ & \phantom{28}0.7\phantom{0} & low coordination number \\
  \hline \multicolumn{4}{c}{Intrinsic errors (other works, VASP-PW91/-PBE/-PBE+U) }\\
  \hline $C_{ij}$ [GPa] \cite{Shang} & \phantom{1}$-2.0\,\%$ & \phantom{2}23\phantom{.00}
& \\
  $\Delta E_{form}$ [kJ/mol] \cite{Lany} & $-13.1\,\%$ &
\phantom{2}15\phantom{.00} & \\
  $\Delta E_{react}$ [kJ/mol] \cite{Hautiererror} & \phantom{1}$-4.8\,\%$ &
\phantom{22}3.2\phantom{0} & chemically dissimilar \\
  $\Delta E_{evol}$ \ [kJ/mol] \cite{USPEX, genetic1, genetic2, genetic3, Ma} & & $\lesssim 1$\phantom{.00} & chemically / structurally dissimilar \\
  \hline \multicolumn{4}{c}{Numerical errors (this work, PBE)} \\
  \hline $\Delta$(PAW)$_\text{(VASP)}$ \ \,[kJ/mol] & & \phantom{22}0.19 & --- \\
  $\Delta$(PAW)$_\text{(GPAW)}$ [kJ/mol] & & \phantom{22}0.32 & --- \\
  \hspace{1.4cm} \vdots & & \phantom{22}\vdots & \\
  \hline \hline
 \end{tabular}
\end{table*}

Based on the quantification of intrinsic errors, a computational recipe has been presented which allows to correct bare DFT-PBE results for the systematic deviation from experiment, and which attaches meaningful error estimates to the obtained predictions. (Tab.~\ref{taberrorproc}). An examination of 31 binary compounds not included in the benchmark set \cite{Haas, Haaserratum} indicate that our analysis carries over to multicomponent crystals. Errors can hence be estimated straightforwardly for PBE predictions already available from literature.

The overall agreement between VASP-PBE and experiment is quite good, but some subsets of elements perform better than others. DFT predictions for magnetic materials and correlation-dominated compounds deviate significantly from experimental values, for example, especially with respect to the cohesive energy. Long-range interaction is another issue. Although some solutions exist to incorporate London dispersion into DFT, such as the \mbox{DFT-D} \cite{Grimme} or vdW-DF2 method \cite{vdwdft}, regular GGAs do not describe dispersion-governed crystal types well. Bulk moduli are found to be consistently underestimated, while predictions for both the volume and the pressure derivative of the bulk modulus are systematically too large. Results for heavy elements are acceptable as long as spin-orbit coupling is added, starting from the 6$p$ block. Based on these observations, some general guidelines have been summarized in Tab.~\ref{tabconcl} as to what categories of materials will not be described well. Some 
classes 
were not 
or only marginally represented in the elemental benchmark set, such as ionic or strongly correlated compounds. For these, a similar study using an extended benchmark set, by including some binary ionic compounds and transition metal oxides, would be useful.

All conclusions with respect to the intrinsic PBE errors can only be universally applicable when it does not matter how the DFT formalism is implemented. Such numerical errors should be much smaller than intrinsic ones. By means of a quality factor $\Delta$, which conveys exactly this information, APW+lo$_\text{(WIEN2k)}$ has been compared to PAW$_\text{(VASP)}$ ($\Delta = 1.9\,$meV/atom) and PAW$_\text{(GPAW)}$ ($\Delta = 3.3\,$meV/atom), both for their recommended sets of atomic potentials. The rms energy distance between equations of state from different methods indeed appears to be an order of magnitude smaller than the gap between theory and experiment (see also Tab.~\ref{tabconcl}). The intrinsic systematic deviations and residual error bars presented in Tab.~\ref{tabconcl} can hence be applied to PBE predictions regardless of the computational approach. This is useful when discussing the implications of DFT results in an experimental context. 

This accuracy review is to be considered as a starting point only. The presented statistical procedure is applicable to other functionals or methods as well. It would be useful to determine the intrinsic systematic deviations and residual error bars for e.g. LDA or hybrid functionals. A comparison to results from high-level many-body techniques would even allow to eliminate the influence of experimental errors. Another extension would be to take into account experimental error bars in the statistical analysis. For the relevant properties of the elemental materials, this requires an extensive literature search for the most accurately known values and their error bars. These error bars are not commonly available in tabulations in the literature and are therefore beyond the scope of this work. With respect to the assessment of numerical errors, we invite both code developers and users to determine the quality factor $\Delta$ for their code as well. It not only guarantees the 
transferability of the intrinsic errors to 
all codes, but also provides a criterion to evaluate the accuracy of a particular DFT approach. A comprehensive list of $\Delta$-factors \cite{molmod} can then serve as a guideline through the maze of available DFT methods.

\begin{acknowledgments}
This work is supported by the Fund for Scientific Research -- Flanders (FWO) and by the Research Board of Ghent University. Stefaan Cottenier acknowledges financial support from OCAS NV by an OCAS-endowed chair at Ghent University. Calculations were carried out using the Stevin Supercomputer Infrastructure at Ghent University, funded by Ghent University, the Hercules Foundation, and the Flemish Government (EWI Department). The authors thank Martijn Marsman for his useful insights on the spin-orbit coupled Pb atom, Fabien Tran for a helpful discussion on zero-point corrections, and Manuel Richter for some interesting comments with respect to scalar-relativistic techniques. 
They acknowledge the implementation by Marcin Du{\l}ak of the code comparison database into ASE \cite{ASE}, and thank him for valuable remarks with respect to the $\Delta$-factor computation script. This work has benefited a lot from remarks and input given by many colleagues at conference presentations, as well as from the extensive and constructive comments provided by three anonymous 
referees.
\end{acknowledgments}

\appendix*

\section*{Appendix: Calculating the $\Delta$-factor}

For a particular compound, the energy difference between the equations of state of WIEN2k $(w)$ and a code under investigation $(c)$ can be evaluated analytically, using the Birch-Murnaghan relation of Eq.~(\ref{eqBM}). Some mathematical manipulations yield
\begin{equation}
 \int_{V_i}^{V_f} \left( E^c(V) - E^w(V) \right)^2 \, \mathrm{d}V = F(V_f) - F(V_i)
\end{equation}
where the primitive function $F(V)$ can be written as a power series in $V^{-1/3}$:
\begin{equation}
 F(V) = \sum \limits_{n=-2}^4 x_n V^{-(2n+1)/3}
\end{equation}
The coefficients $x_n$ are given by
\begin{equation}
 x_n = - \frac{3}{2n+1} \sum \limits_{i+j=n+2} (a_i^{c} - a_i^{w}) (a_j^{c} - a_j^{w}) \label{eqxn}
\end{equation}
with $i,j \in \{0, 1, 2, 3\}$. Then $x_{-1}$ for example becomes
\begin{equation}
 x_{-1} = 6 ( a_1^c - a_1^w) (a_0^c - a_0^w)
\end{equation}
The constants $a_i$ are the coefficients of the Birch-Murnaghan equation in its polynomial form:
\begin{align}
 a_3 &= \frac{9 V_0^3 B_0}{16} (B_1 - 4) \\
 a_2 &= \frac{9 V_0^{7/3} B_0}{16} (14 - 3 B_1) \\
 a_1 &= \frac{9 V_0^{5/3} B_0}{16} (3 B_1 - 16) \\
 a_0 &= \frac{9 V_0 B_0}{16} (6 - B_1)
\end{align}
When evaluating Eq.~(\ref{eqxn}), $a_i^c$ stands for the coefficient of the code under test, while $a_i^w$ means it corresponds to the WIEN2k equation of state.

\bibliography{bibliografie3}

\end{document}